\documentclass[pdflatex,sn-mathphys-ay]{sn-jnl}

\usepackage{lmodern}
\usepackage{anyfontsize}
\newcommand\aap{A\&A}                
\newcommand\ao{Appl. Opt.}           
\newcommand\apjl{ApJ}                
\newcommand\apjs{ApJS}               

\usepackage{graphicx}%
\usepackage{multirow}%
\usepackage{amsmath,amssymb,amsfonts}%
\usepackage{amsthm}%
\usepackage{mathrsfs}%
\usepackage[title]{appendix}%
\usepackage{xcolor}%
\usepackage{textcomp}%
\usepackage{manyfoot}%
\usepackage{booktabs}%
\usepackage{algorithm}%
\usepackage{algorithmicx}%
\usepackage{algpseudocode}%
\usepackage{listings}%
\usepackage{amsmath}
\usepackage{multirow}
\usepackage{amssymb}
\usepackage{pifont}
\usepackage[numbers]{natbib}
\newcommand{\cmark}{\ding{51}}%
\newcommand{\xmark}{\ding{55}}%
\usepackage{mathrsfs}

\theoremstyle{thmstyleone}%
\theoremstyle{thmstyletwo}%

\theoremstyle{thmstylethree}%
\jyear{2024}%
\begin{document}

\title[Camelidae on BOAT]{Camelidae on BOAT: observation of a second spectral component in GRB 221009A}


\author*[1,2]{\fnm{Biswajit} \sur{Banerjee}}\email{biswajit.banerjee@gssi.it}

\author[1,2]{\fnm{Samanta} \sur{Macera}}
\equalcont{These authors contributed equally to this work.}

\author[1,2]{\fnm{Alessio Ludovico} \sur{De Santis}}
\equalcont{These authors contributed equally to this work.}

\author[1,2]{\fnm{Alessio} \sur{Mei}}
\equalcont{These authors contributed equally to this work.}

\author[1,2]{\fnm{Jacopo} \sur{Tissino}}
\equalcont{These authors contributed equally to this work.}

\author*[1,2]{\fnm{Gor} \sur{Oganesyan}}\email{gor.oganesyan@gssi.it}

\author[3]{\fnm{Dmitry D.} \sur{Frederiks}}

\author[3]{\fnm{Alexandra L.} \sur{Lysenko}}

\author[3]{\fnm{Dmitry S.} \sur{Svinkin}}

\author[4,3]{\fnm{Anastasia E.} \sur{Tsvetkova}}

\author[1,2]{\fnm{Marica} \sur{Branchesi}}



\affil[1]{ \orgname{Gran Sasso Science Institute}, \orgaddress{\street{Viale F. Crispi 7}, \city{L'Aquila (AQ)}, \postcode{I-67100}, \country{Italy}}}

\affil[2]{ \orgname{INFN - Laboratori Nazionali del Gran Sasso}, \orgaddress{ \city{L'Aquila (AQ)}, \postcode{I-67100}, \country{Italy}}}

\affil[3]{\orgname{Ioffe Institute}, \orgaddress{\street{26 Politekhnicheskaya}, \city{St Petersburg}, \postcode{194021}, \country{Russia}}}

\affil[4]{\orgdiv{Department of Physics}, \orgname{University of Cagliari}, \orgaddress{\street{SP Monserrato-Sestu}, \city{Cagliari}, \postcode{09042}, \state{Cagliari}, \country{Italy}}}


\abstract{
\noindent  
Observing and understanding the origin of the very-high-energy (VHE) spectral component in gamma-ray bursts (GRBs) has been challenging because of the lack of sensitivity in MeV-GeV observations, so far. The majestic GRB 221009A, known as the brightest of all times (BOAT), offers a unique opportunity to identify spectral components during the prompt and early afterglow phases and probe their origin. Analyzing simultaneous observations spanning from keV to TeV energies, we identified two distinct spectral components during the initial 20 minutes of the burst. 
The second spectral component peaks between $10-300$ GeV, and the bolometric fluence (10 MeV-10 TeV) is estimated to be greater than 2$\times10^{-3}$ erg/ cm$^{2}$. 
Performing broad-band spectral modeling, we provide constraints on the magnetic field and the energies of electrons accelerated in the external relativistic shock. We interpret the VHE component as an afterglow emission that is affected by luminous prompt MeV radiation at early times.}




\maketitle


In 2019, the two Imaging Atmospheric Cherenkov Telescopes (IACTs) Major Atmospheric Gamma-ray Cherenkov Telescopes (MAGIC) and the High Energy Stereoscopic System (H.E.S.S.) discovered TeV emission from gamma-ray bursts. GRB 190114C \cite{GRB190114C_MAGIC}, located at redshift $z=0.46$, was detected by MAGIC approximately 62\,s after the Gamma ray Burst Monitor on board the Fermi satellite (Fermi/GBM) trigger time. This pivotal observation of GRB 190114C allowed an analysis of the time-resolved very-high-energy gamma-ray (VHE; E $>$ 100\,GeV) spectrum across multiple epochs, extending well beyond 2000\,s. Spectral analysis, using concurrent data, was reported to be consistent with synchrotron self-Compton radiation from the electrons accelerated in the forward shock \cite[see][ for a review]{Miceli_2022}. 
The H.E.S.S. collaboration detected the VHE emission of GRB 180720B at redshift $z = 0.653$ \cite{Abdalla_2019} 
occurring 10\,ks after the Fermi/GBM trigger time. Later, H.E.S.S also detected GRB 190829A \cite{GRB190829A}. This observation was used to perform a joint spectral analysis with Swift/XRT during two time periods: after about 1) 4- 8\,hr and 2) 27- 32\,hrs. The joint analysis reveals that a single emission component could in principle account for the broadband spectra. A separate study by \cite{Salafia_2022} has shown that the SSC model is capable of producing the observed spectral energy distribution for this GRB. \par

To date, several GRBs, for example, GRB 201216C \cite{Abe_2023}, GRB 201015A \cite{Nava:2021yfk} were discovered by IACTs, such as MAGIC and H.E.S.S. However, the inability to impose constraints in the high-energy gamma-ray spectrum (50 MeV-100 GeV) prevents an appropriate evaluation of the emission mechanism. 
For GRB 190114C and GRB 180720B, confirmation of a secondary component that exceeded the MeV spectrum was unattainable \cite{GRB190114C_MAGIC, Abdalla_2019, GRB190829A}, as the minimal significance of detection in the GeV spectrum restricted the spectral analysis.
Consequently, the lack of detection of a significant secondary component limits the ability to investigate the microphysical parameters of the shock.\par

On 9 October 2022, the gamma-ray burst GRB 221009A was detected by both ground-based and space-borne observatories and subsequently identified as one of the most energetic bursts on record \cite{Lesage_2023, Frederiks_2023} at redshift 0.15 \cite{Malesani2023}. This intense burst was observed during its initial emission phase by Fermi/GBM and Konus-$Wind$, leading to its identification as having the highest prompt fluence and peak flux \cite{Lesage_2023, Frederiks_2023, Burns_2023}. 
The intense fluence saturated GBM during specific phases, termed bad time intervals (BTI), BTI-1: 218.5-277.9\,s and BTI-2: 507.3-514.4\,s post the GBM trigger time, which were marked as unusable intervals for the data analysis \cite{Lesage_2023}.
During the initial phase up to 247\,s, around the peak of the brightest pulse, only Konus-$Wind$ 
was able to recover the prompt-emission spectrum of GRB 221009A \cite{Frederiks_2023}. 
Most interestingly, LHASSO observed GRB 221009A \cite{LHAASO2023} starting from the onset of the burst, thanks to its wide FoV and high duty cycle \cite{Cao_2024}. The time-resolved spectral analysis covering the energy band 0.3-5\,TeV, revealed a VHE emission up to approximately one hour after the trigger. In this study, we consider the observation periods of LHAASO as reference time-bins.  \par

\begin{figure*}[ht]\caption{Illustration of the data availability in each time bins (time stamps indicated in vertical axis on the left and numbered bins on the right) for each instrument (see Table \ref{tab:MWLdata}). Areas marked with slashes indicate bins impacted by pile-up in both Fermi/GBM and Fermi/LAT. A simple double-hump spectrum is shown to indicate the preliminary peak positions of the two components seen in GRB 221009A. The selection of time bins is based on the presence of MWL data spanning keV to TeV energies. Six bins (BIN-2, -6, -7, -8, -12, and -14) are chosen for time-resolved MWL spectral analysis.}
\centering 
\includegraphics[width=1.1\textwidth]{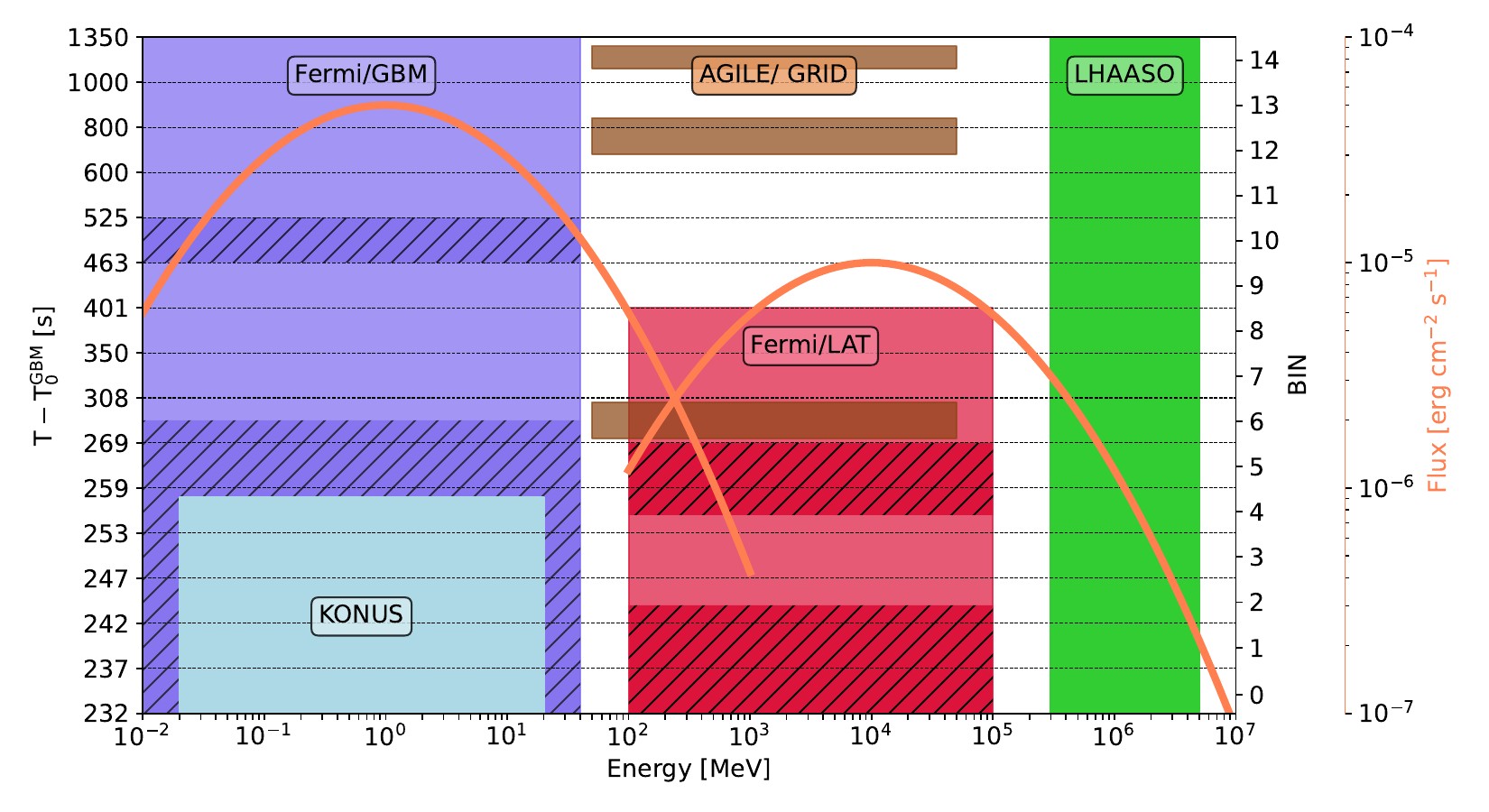}
\label{fig:main_LC1}
\end{figure*}

\begin{figure*}\caption{Spectra obtained with the available observations (starting from GBM trigger time) from different instruments in six selected time bins on the basis of the presence of GeV detections by LAT and/or AGILE complemented with intrinsic spectra from LHAASO. Additional data from GBM (NaI4, NaI8, B1) and KONUS in these bins are shown, where available. Overall, the data show the existence of two separate components. The blue band, which spans energies from 10 keV to 100 GeV, represents the bet-fit model for MeV-range data (see Table \ref{tab:GBM_SEDmodel}). The green band, covering 10 MeV to 10 TeV, indicates the optimally fitted log-parabola model for GeV-TeV range energies (see the Methods for details).}
\centering 
    \includegraphics[width=0.48\textwidth]{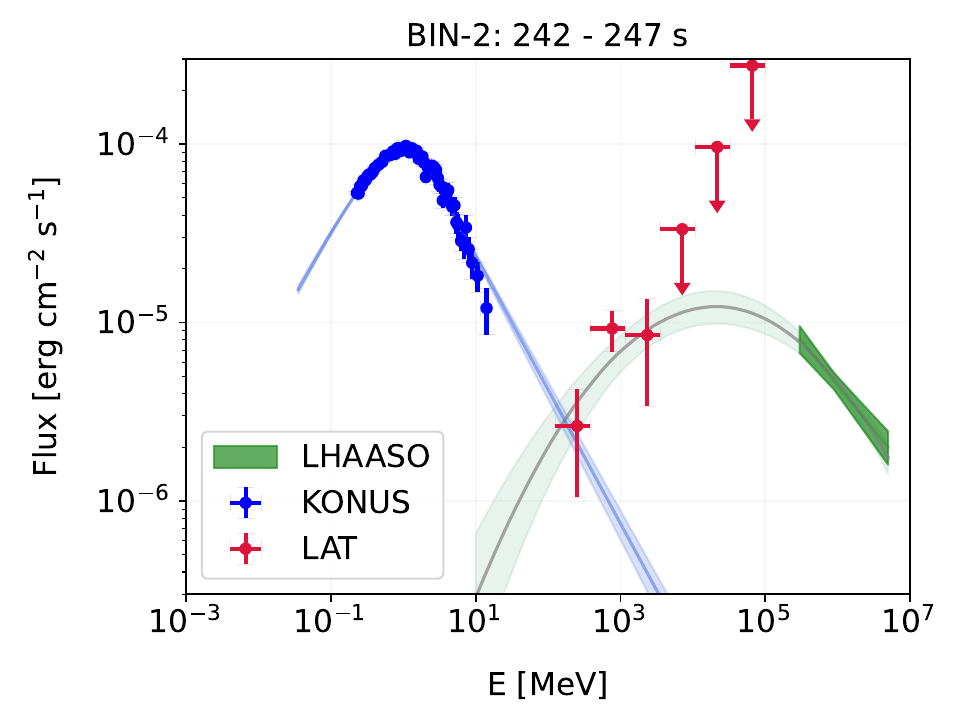}
    \includegraphics[width=0.48\textwidth]{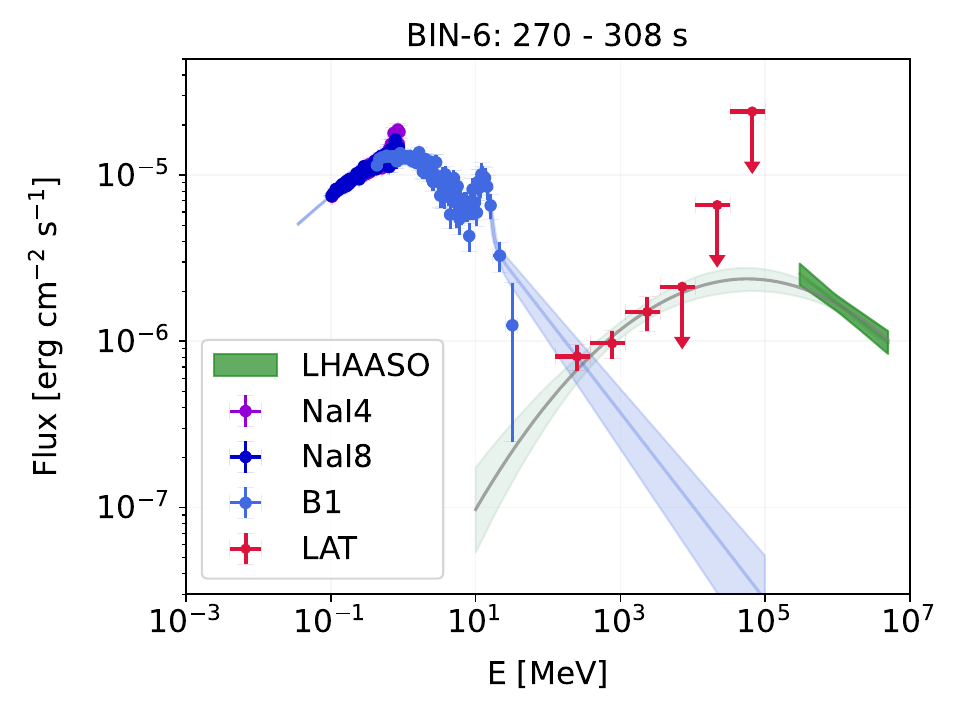}
    \includegraphics[width=0.48\textwidth]{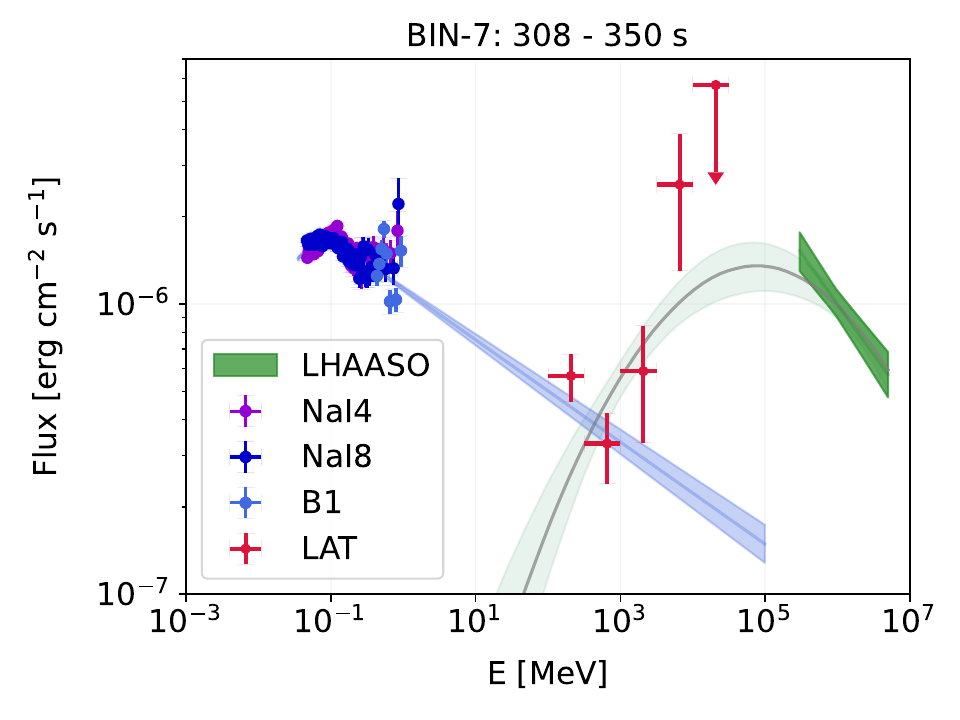}
    \includegraphics[width=0.48\textwidth]{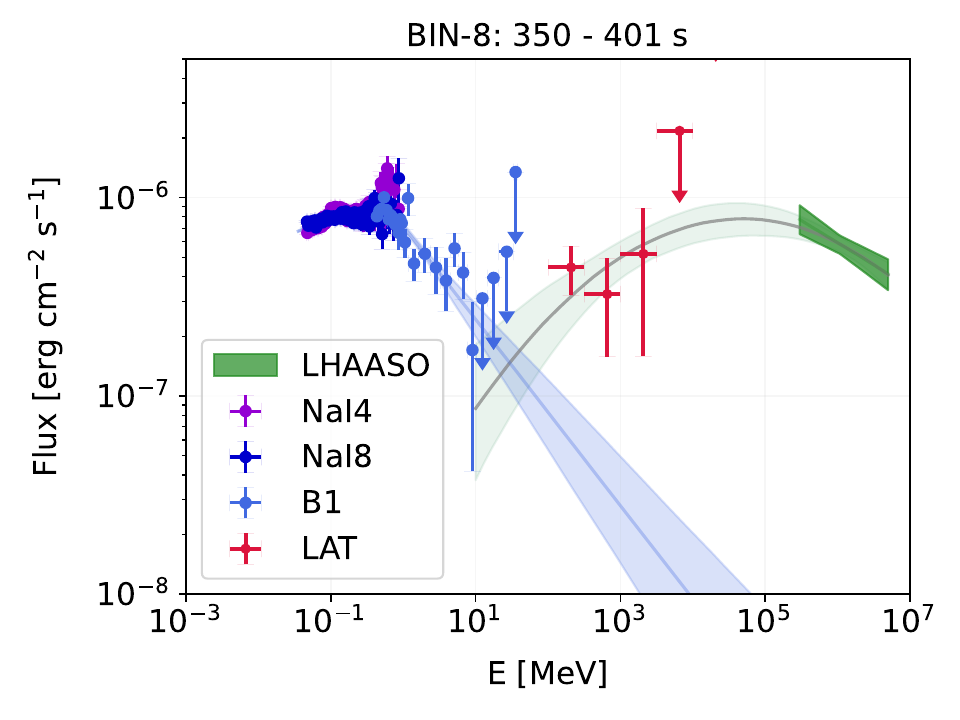}
    \includegraphics[width=0.48\textwidth]{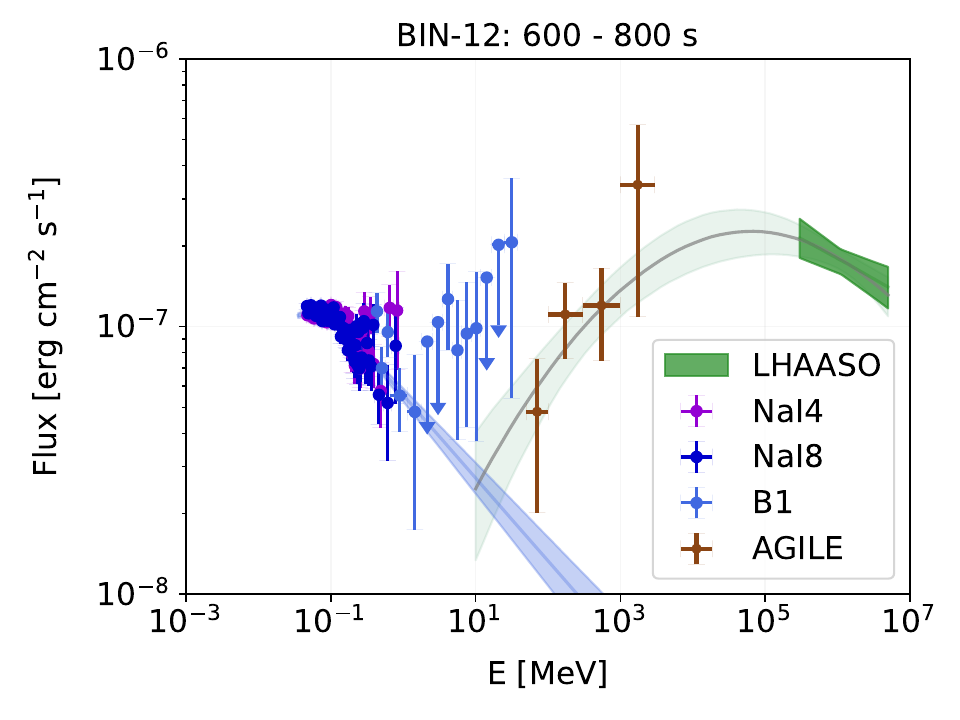}
    \includegraphics[width=0.48\textwidth]{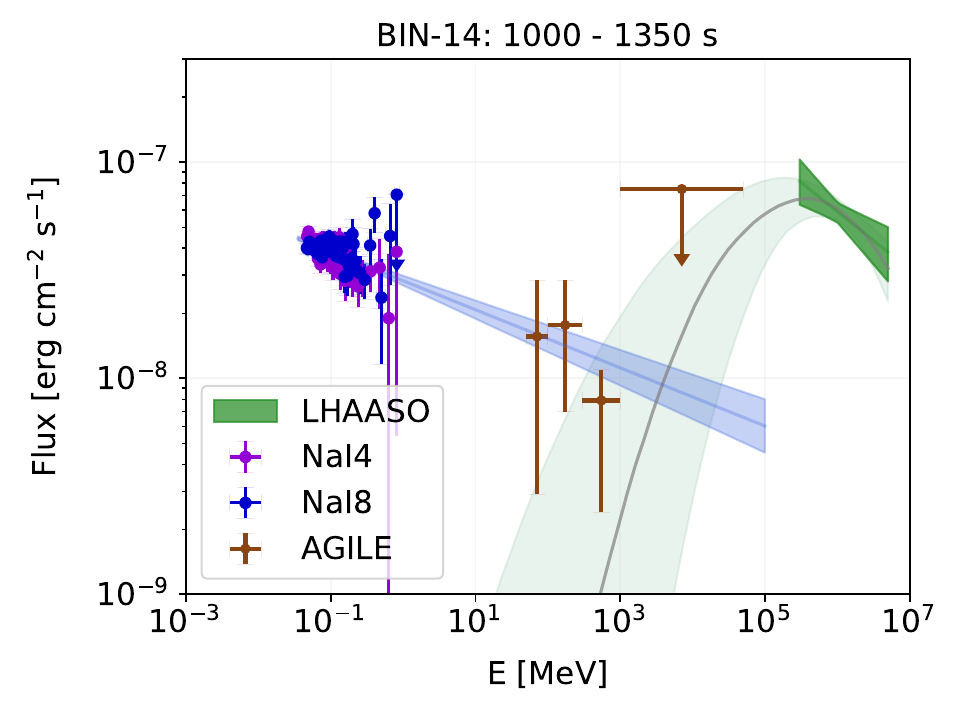}
	\label{fig:SED_Fig2}
\end{figure*}

In the high-energy gamma-rays from 50 MeV and up to 100 GeV, the Large Area Telescope (LAT) on board the Fermi satellite observed the burst from the onset until the GRB moved out of LAT's field of view (FoV) around 400\,s. Due to the high fluence in hard X-rays, the LAT data experienced significant pile-up during the time interval 219-244\,s and 255-270\,s, making the GeV data during these times unrecoverable. Data from 244-255\,s and 270-280\,s, while still affected by pile-up, could be analyzed using a specific method described in \cite{Bissaldi:2023gi}.
AGILE recorded observations of GRB 221009A \cite{Tavani_2023}. Specifically, it captured data during two separate intervals, 684.0-834.0\,s and 1129.0-1279.0\,s, which extend beyond the LAT observation period (after 400\,s) and closely align in timing \cite{Tavani_2023} with the intervals reported by LHAASO. 
Referring to time intervals prompted by LHAASO observations (see Table \ref{tab:MWLdata}, spanning from BIN-0 to -14), we select specific time bins based on the availability of quasi-simultaneous data from various instruments. The time bins and available data are shown in Fig. \ref{fig:main_LC1}. The bins of interest, specifically BIN-2: 242.0-247.0, BIN-3: 247.0-253.0, BIN-6: 270.0-308.0, BIN-7: 308.0-350.0, BIN-8: 350.0-401.0, BIN-12: 600.0-800.0, and BIN-14: 1000.0-1350.0 provide comprehensive data coverage from keV to TeV energies. The multiwavelength data for these bins are shown in Fig. \ref{fig:SED_Fig2}. Using these time bins, we construct the multiwavelength spectrum (covering energies from keV to TeV) using the concurrent data available from keV-MeV (lower-energy band covered with Fermi/GBM and KONUS) to GeV energies (Fermi/LAT and AGILE) along with the TeV data from LHAASO. \par

Observations from GBM and KONUS in the MeV energies are fitted with a standard Band function for the prompt emission spectra \cite{1993ApJ...413..281B}, with the exception of BIN-6, which in addition to a Band-function requires a distinct Gaussian line emission (the Ravasio line), as discovered in \cite{ravasio2023bright}. The optimal model parameters are detailed in Table \ref{tab:GBM_SEDmodel}. In each time bin, the model is then extended up to 100 GeV. 

Across all the temporal bins mentioned above, the soft spectral slope of the emission at MeV energies (see Table \ref{tab:GBM_SEDmodel}, third column) and the presence of the GeV and TeV component exceeding the best-fit model from the keV-MeV band extended up to 100 GeV clearly indicate the emergence of the second spectral component. This is particularly evident in the earlier bins (such as BIN-2, BIN-6), where the source exhibits high-flux states across all energy bands. Furthermore, (for example, BIN-2) a slope change in the GeV data compared to the keV-MeV data has been observed, further corroborating the presence of the second component. For the final time-bin, BIN-14, the Fermi/ GBM data extending up to 100 GeV (illustrated in the blue band in Fig. \ref{fig:SED_Fig2}) align with the flux and slope of the GeV data obtained with AGILE. Consequently, the extended model, supported by AGILE's observations, predicts a flux in the VHE gamma-rays that is a factor of 10 lower, as observed by LHAASO in the 0.3-5 TeV range, confirming the existence of the second component even at late times beyond 1000 seconds.\par

In addition, we fit a log-parabola model to characterize the GeV (observed by Fermi/LAT and AGILE) and TeV (observed by LHAASO) data.  For the six time intervals, we determine the peak position, peak flux, and bolometric flux in the energy band of 10 MeV to 10 TeV (see Methods for more details). We find that
the peak position does not show significant variation and ranges between 10\,GeV- 300\,GeV (see Fig. \ref{fig:LogParModel2} in Methods). Instead, the peak positions estimated at later times (after 1000\,s) are not well constrained due to the data quality in the GeV band. The bolometric flux in the energy band of 10 MeV-10 TeV roughly traces the light curve of LHAASO (see Fig. \ref{fig:VHE_FBol} in the Methods).  

\begin{figure*}\caption{The light curve of GRB 221009A observed by several instruments including LHAASO in the 0.3-5\,TeV, Fermi/LAT and AGILE in the 0.1-3\,GeV, GBM and KONUS in 0.04-40 MeV range. 
The details of the analysis of the Fermi LAT data are discussed in Methods. The AGILE data, obtained from \cite{Tavani_2023}, have been rescaled in the energy band 0.1-3\,GeV (detailed in the Methods). The light curve was plotted with the choice of a trigger time of T$^{'}_{0}$ = 177\,s after the GBM trigger time (T$^{\rm GBM}_{0}$) which marks the arrival of the first MeV pulse as reported by \cite{Frederiks_2023}. The light curves in the energy bands mentioned above tend to show a similar temporal decay (with a similar energy flux) about after (about) 450\,s which translates to about 630\,s from the GBM trigger time. This time represents the onset of the afterglow emission. The two phases (BTI-1 and BTI-2) marked with a red-shadded region denote the BTI intervals of Fermi/GBM.
}\label{fig:GeVandTeV}
\centering 
    \includegraphics[width=\linewidth]{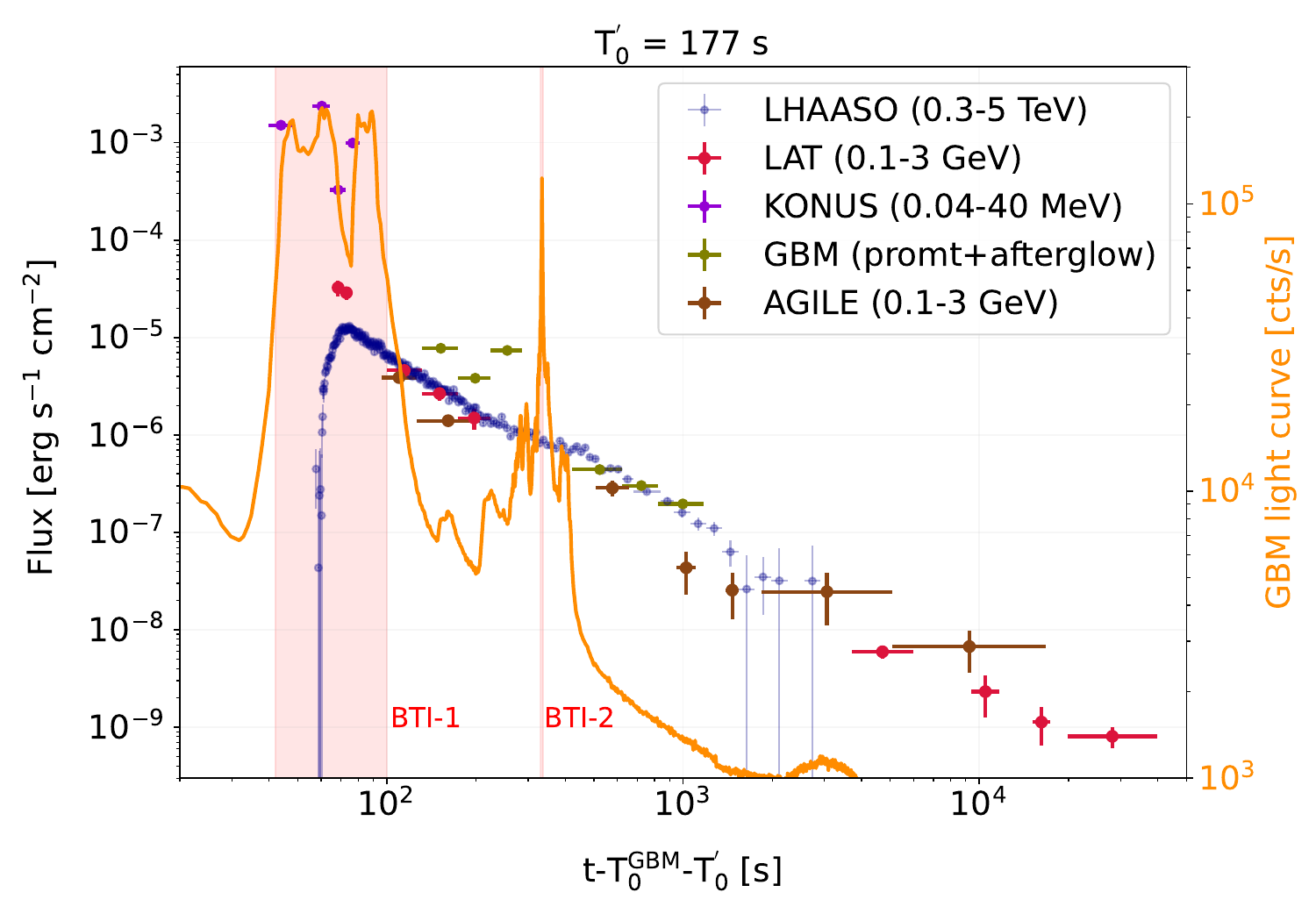}
\end{figure*}

To evaluate the underlying mechanism responsible for the broadband emission during the first 20 minutes, we constructed the light curve across the following energy bands: 1) 0.3-5\,TeV (LHAASO), 2) 0.1-3\,GeV (LAT and AGILE data when available) and 3) 0.04-40\,MeV (GBM and KONUS). The MWL light curve is shown in Fig. \ref{fig:GeVandTeV}. The choice of the 0.04-40 MeV energy band is because it is covered by both Fermi/GBM and KONUS, whereas 0.1-3 GeV is covered by both Fermi/LAT and AGILE. Furthermore, we consider a start time (T$^{'}_{0}$) of 177\,s, corresponding to the onset of the burst as observed by KONUS \cite{Frederiks_2023}. We observe that the variability in the light curve becomes less pronounced starting from BIN-12, aligning with the period when the variable emission in GBM ceases and shows a steady decline. This is indicative of the the afterglow emission.\par

\begin{figure*}\caption{The multi-wavelength spectral energy distribution (SED) of GRB 221009A at the latest epoch (1000-1350 s). The SED is modelled with SSC. The 65\% confinement of the best fit model is presented in yellow shadded region. The posterior distribution of parameters is shown in Fig. \ref{fig:cornerSSC1} and the parameters are listed in Table~\ref{tab:bestfitmodel}.}
\label{fig:BIN14_model}
\centering 
    \includegraphics[width=0.8\linewidth]{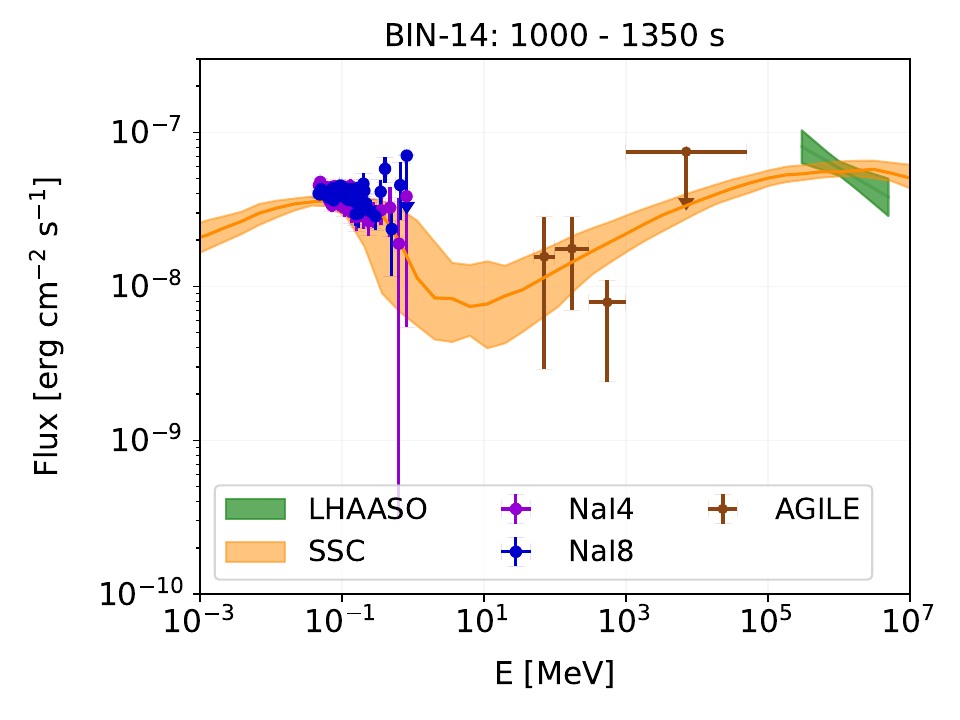}
\end{figure*}

The MeV spectral component of GRB 221009A is dominated by the prompt emission for the first $\sim 500$\,s as inferred from its variability. Given the absence of variability in the TeV light curve as observed by LHAASO, the MeV and GeV-TeV spectral components are not produced from the same emission region of the early epochs ($< 500$ s). In contrast, the late MeV emission ($>600$ s) is temporally featurless and can be assigned to the afterglow emission component, as also considered by other authors \cite{ZhangH2023,Lesage_2023,An2023,Zheng2024}. We thus consider the latest ($1000-1350$ s) MeV-TeV spectrum to rise from the synchrotron and synchrotron self-Compton (SSC) radiation of the forward shock-accelerated electrons of the circum-burst medium. The 40 keV - 1 MeV spectrum is best fitted by the power-law model with the photon index of $-2.14 \pm 0.02$. The $95 \%$ upper limit of the peak energy is 53 keV. The 100 MeV - 5 TeV is described by a log-parabola model with the peak energy of $\approx 3 \times 10^{5}$ MeV. In the SSC scenario, and assuming the Thomson regime, it would require electrons to have energies $\ge 0.3 \, GeV$. 

In this scenario, we access the microphysical parameters of the external shock modeling the joint keV-TeV spectrum of the afterglow emission at the latest epoch (1000-1350 s) with the SSC radiation model (see Fig. \ref{fig:BIN14_model} and the Methods). The data is best described by the soft electron population ($dN/dE_{e} \propto E_{e}^{-2.7}$) in the magnetic field of $< \, 5 \times 10^{-2}$ G. We find that the energies of the injected electrons span at least 3 orders of magnitude with the minimum energy $< \, 2$ GeV. In the standard afterglow model \cite{Meszaros1997,Sari1998,Panaitescu2000}, it corresponds to low values for equipartition parameters for non-thermal electrons ($\epsilon_e < 0.01$) and for the magnetic field ($\epsilon_B < 2 \times 10^{-5} n_{-1}^{-1}$), where n is the density of the circumburst medium. For the fiducial value of $n=0.1 \, cm^{-3}$, the magnetization parameter is $\sigma \, < \, 1 B^{2}_{-1.3} n_{-1}^{-1}$. To reach an efficient acceleration in the relativistic shock, one would require a low magnetization of $\sigma \le 10^{-3}$ in the pair plasma or $\sigma \le 10^{-5}$ in the electron-ion plasma \cite{Sironi2013}. This requires the magnetic field to be $B \le 2 \times 10^{-3}$ G or $B \le 2 \times 10^{-4}$ G, within the posterior distribution of B (Fig. \ref{fig:cornerSSC1}). 

\begin{figure*}\caption{Predicted SEDs for the early epochs: at 270-308 s (BIN-6) and 600-800 s (BIN-12). The SEDs are estimated from the posterior distribution of SSC model for the latest epoch at 1000-1350 s (BIN-14), with the self-similar dynamics of the relativistic blast wave in the homogenous (dark orange) circumburst medium. Note that these are not the fits to the data but predictions of the SEDs from the scaling of microphysical parameters according to the expected dynamics of the forward shock.}
\label{fig:Extended_model}
\centering 
    \includegraphics[width=0.48\linewidth]{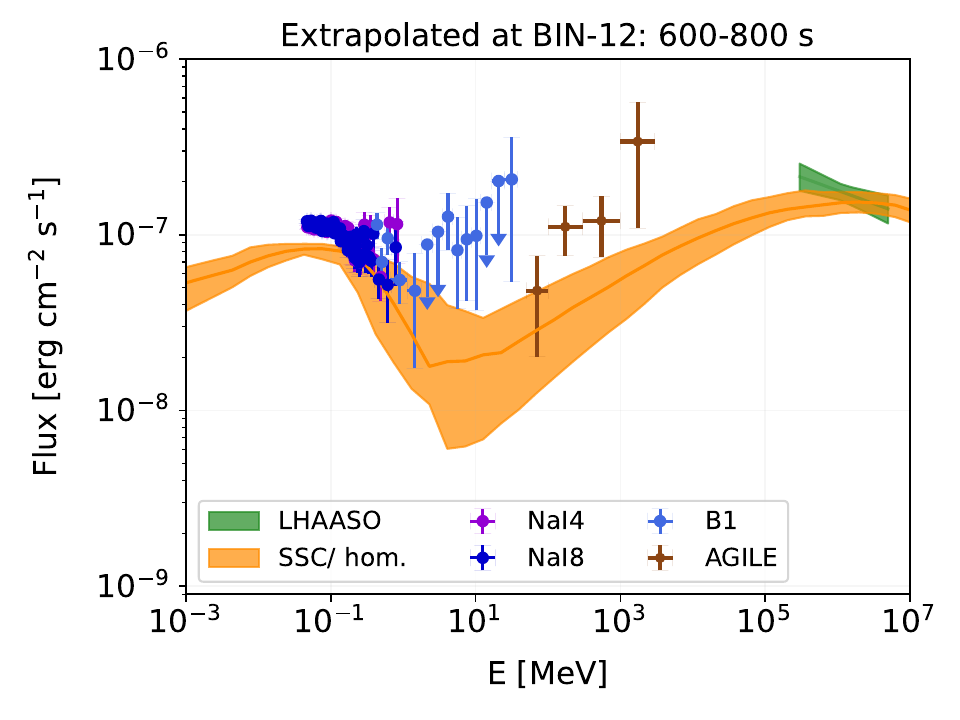}
    \includegraphics[width=0.48\linewidth]{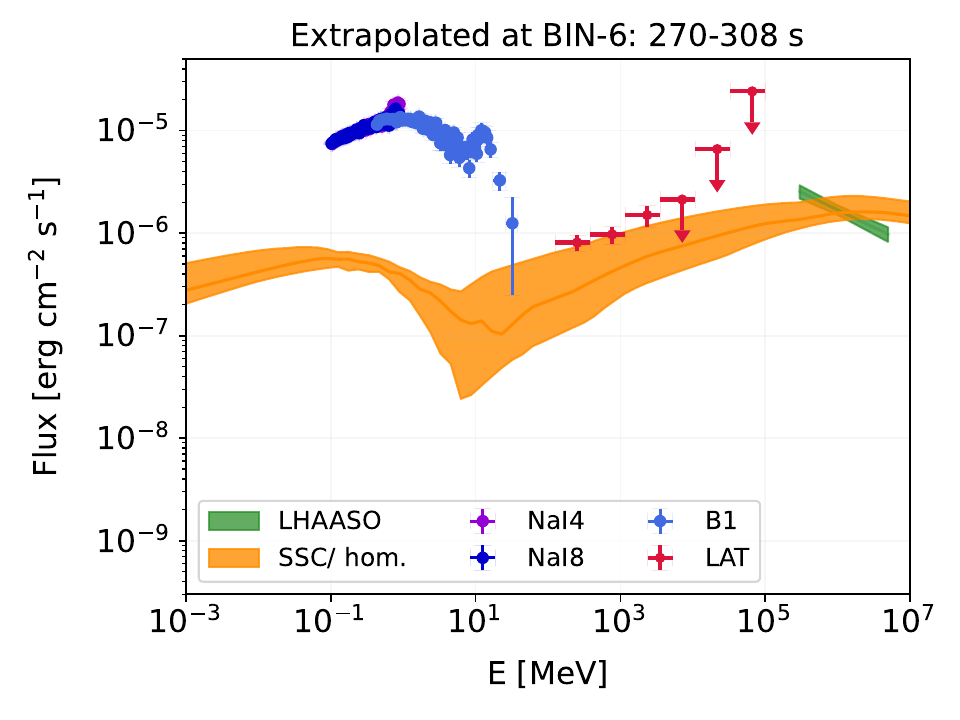}
\end{figure*}

\begin{figure*}\caption{
The light curve of LHAASO (0.3-5 TeV; blue points) together with the expected light curve of the afterglow emission (yellow line) in homogeneous circumburst medium. The model is obtained using the median value of the microphysical parameters of the SSC model for the latest epoch at 1000-1350 s (BIN-14), with the self-similar dynamics of the relativistic blast wave. Note that this is not a fit to the data, but the prediction of the afterglow radiation from the scaling of microphysical parameters according to the expected dynamics of the forward shock.}
\label{fig:SSC_LC}
\centering 
    \includegraphics[width=0.9\linewidth]{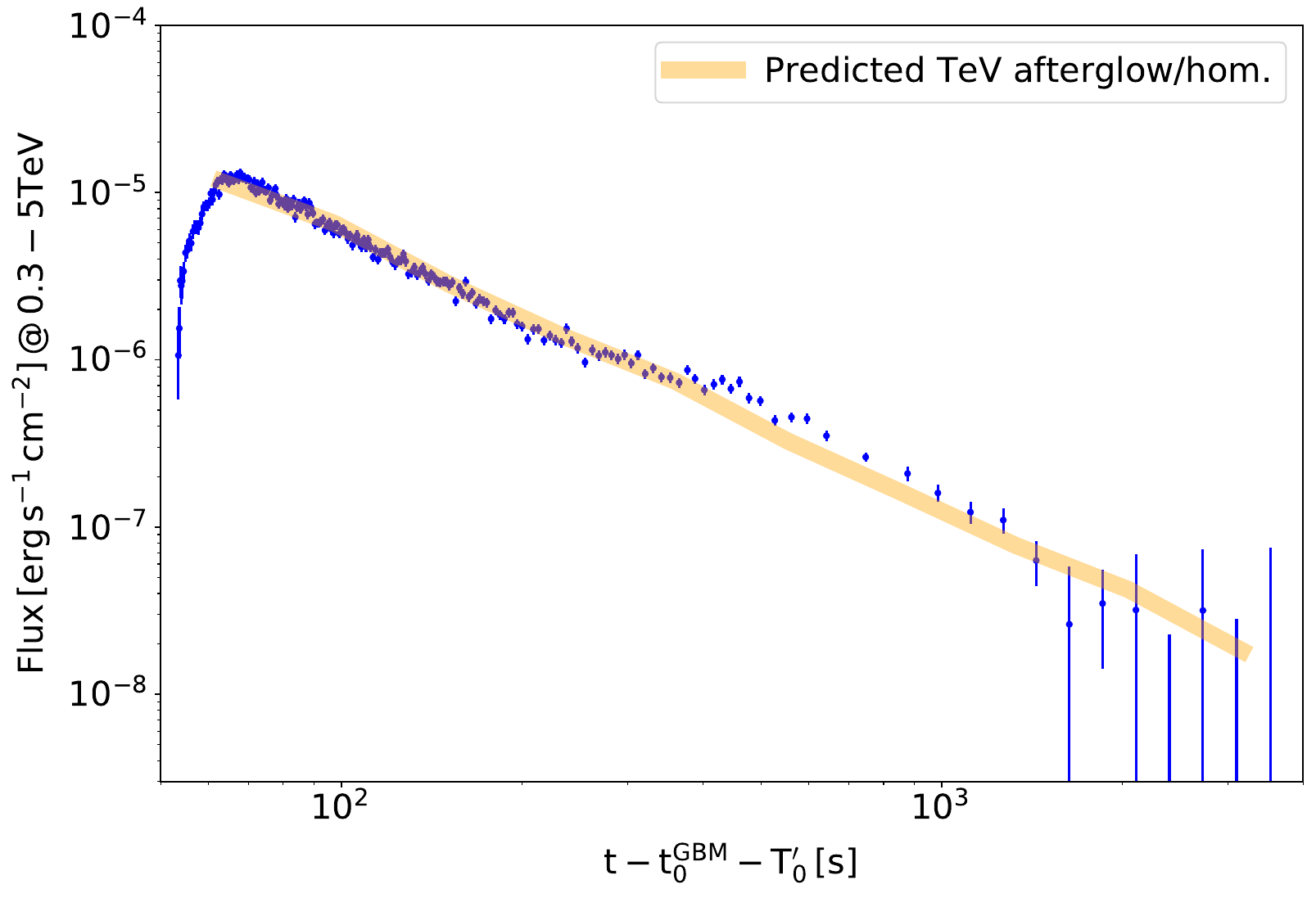}
\end{figure*}

Regardless of the exact value of the density of the circumburst medium and the total kinetic energy of the jet, it is possible to estimate the broad-band spectra of afterglow emission at earlier times due to the self-similar nature of the relativistic blast wave dynamics in the cold medium \cite{BM1976}. To predict the expected keV-TeV afterglow emission at early epochs, we assume that the equipartition parameters and the index of the accelerated electrons do not vary with time. Under these assumptions, we derive the keV-MeV spectra of the SSC radiation of the afterglow at earlier epochs (see Fig. \ref{fig:Extended_model} and the Methods). In the homogeneous circumburst case, the keV-TeV spectra are within the predicted model. We also compute the expected 0.3-5 TeV light curve in the SSC scenario (see Fig. \ref{fig:SSC_LC}). The observed LHAASO light curve is in agreement with this simple model with slight deviations. However, one could notice that the observed TeV spectra are slightly softer compared to those expected in the standard SSC model for afterglow emission. Fig. \ref{fig:Extended_model} shows that the AGILE and Fermi/LAT flux are slightly in excess with respect to the SSC predictions. We interpret these features as due to external supply of prompt emission photons to electrons accelerated in the forward shock (see Methods for details).

The broad energy coverage (40 keV - 5 TeV) of the GRB 221009A prompt and the afterglow spectra have allowed us to distinguish the VHE spectral component for the first time. In the six time-resolved spectra spanning over the first 20 minutes after the GRB trigger, we found the VHE component to peak at 10-100 GeV. In later epochs ($>$ 600 s), we obtained a two-component SED of the afterglow radiation. Assuming the leptonic origin of the afterglow radiation, we have modeled the keV-TeV spectrum at the latest epoch (1000-1350 s) by the SSC radiation model of the power-law distributed electrons. This data is best described by a distribution of electrons with energies ranging from $<$ 2 GeV to 0.5 TeV and have a soft distribution ($dN/dE_{e} \propto E_{e}^{-2.7}$). We used the results of modeling of single-epoch data to predict early keV-TeV data. In the simplest model assumptions of constant equipartition parameters and the expansion of the relativistic blast wave in the homogeneous medium, we obtained good agreement for the 0.3-5 TeV light curve (Fig. \ref{fig:SSC_LC}) and the SEDs (Fig. \ref{fig:Extended_model}). We have also discussed the effects of MeV photons of the prompt emission in pair production/External Inverse Compton (EIC) cooling with the photons/electrons produced in the forward shock. The prompt emission photons suppress (pair production) and amplify (EIC) the observed VHE radiation (with a delay of $\sim$ 4t). These effects are more pronounced in the observed spectra rather than in the overall TeV light curve of the afterglow emission from the onset of deceleration.

\newpage

\clearpage

\section*{Methods}\label{sec:methods}
\begin{table*}\caption{The \textbf{intrinsic spectral parameters} for LHAASO. The rebinning of the time-resolved LHAASO light curve (fitted with a zero-order polynomial) is described in Methods. The spectral parameters are obtained from \cite{LHAASO2023}. The intrinsic parameters mentioned in this table are further used to produce the very-high-energy gamma-ray "bow-tie" plots in Fig. \ref{fig:SED_Fig2}.} 
    \label{tab:LHAASO_rebinned_table}
    \centering
    \begin{tabular}{ l c c c } \hline
Bins & Time [s] & Flux [$\times$ 10$^{-7}$\,erg\,cm$^{-2}$\,s$^{-1}$] & Index\\ \hline
BIN-0 &232.0-237.0 &64.52$\pm$7.10 & 2.42$\pm$0.08\\ 
BIN-1 &237.0-242.0 &108.35$\pm$11.77 & 2.46$\pm$0.06\\
BIN-2 &242.0-247.0 &122.65$\pm$13.30 & 2.51$\pm$0.07\\ 
BIN-3 &247.0-253.0 &105.33$\pm$11.41 & 2.30$\pm$0.06\\ 
BIN-4 &253.0-259.0 &90.12$\pm$9.80 & 2.44$\pm$0.06\\ 
BIN-5 &259.0-270.0 &76.04$\pm$8.22 & 2.28$\pm$0.05\\ 
BIN-6 &270.0-308.0 &46.08$\pm$4.96 & 2.36$\pm$0.04\\ 
BIN-7 &308.0-350.0 &27.36$\pm$2.95 & 2.37$\pm$0.05\\ 
BIN-8 &350.0-400.0 &16.15$\pm$1.75 & 2.27$\pm$0.05\\ 
BIN-9 &400.0-460.0 &11.55$\pm$1.26 & 2.23$\pm$0.06\\ 
BIN-10 &460.0-525.0 &8.97$\pm$0.99 & 2.32$\pm$0.07\\ 
BIN-11 &525.0-600.0 &7.35$\pm$0.82 & 2.22$\pm$0.06\\ 
BIN-12 &600.0-800.0 &4.91$\pm$0.54 & 2.15$\pm$0.06\\ 
BIN-13 &800.0-1000.0 &2.62$\pm$0.32 & 2.23$\pm$0.08\\ 
BIN-14 &1000.0-1350.0 &1.61$\pm$0.20 & 2.27$\pm$0.12\\ \hline
    \end{tabular}
\end{table*}
\begin{figure*}\caption{The extraction of the energy flux observed by LHAASO from publicly available data for the periods reported with spectral index. In the required time segments (named time-bins: see Table \ref{tab:MWLdata}) the light curve of LHAASO has been fitted with a constant (pol-0) which represents the flux of GRB 221009A for that period. The blue markers in the top panel report the rebinned LHAASO flux for that period. Moreover, a systematic undertainity of 10.7\% has been considered as mentioned in \cite{LHAASO2023}. The bottom panel reports the intrinsic spectral index for the same periods.}
\centering 
    \includegraphics[width=\linewidth]{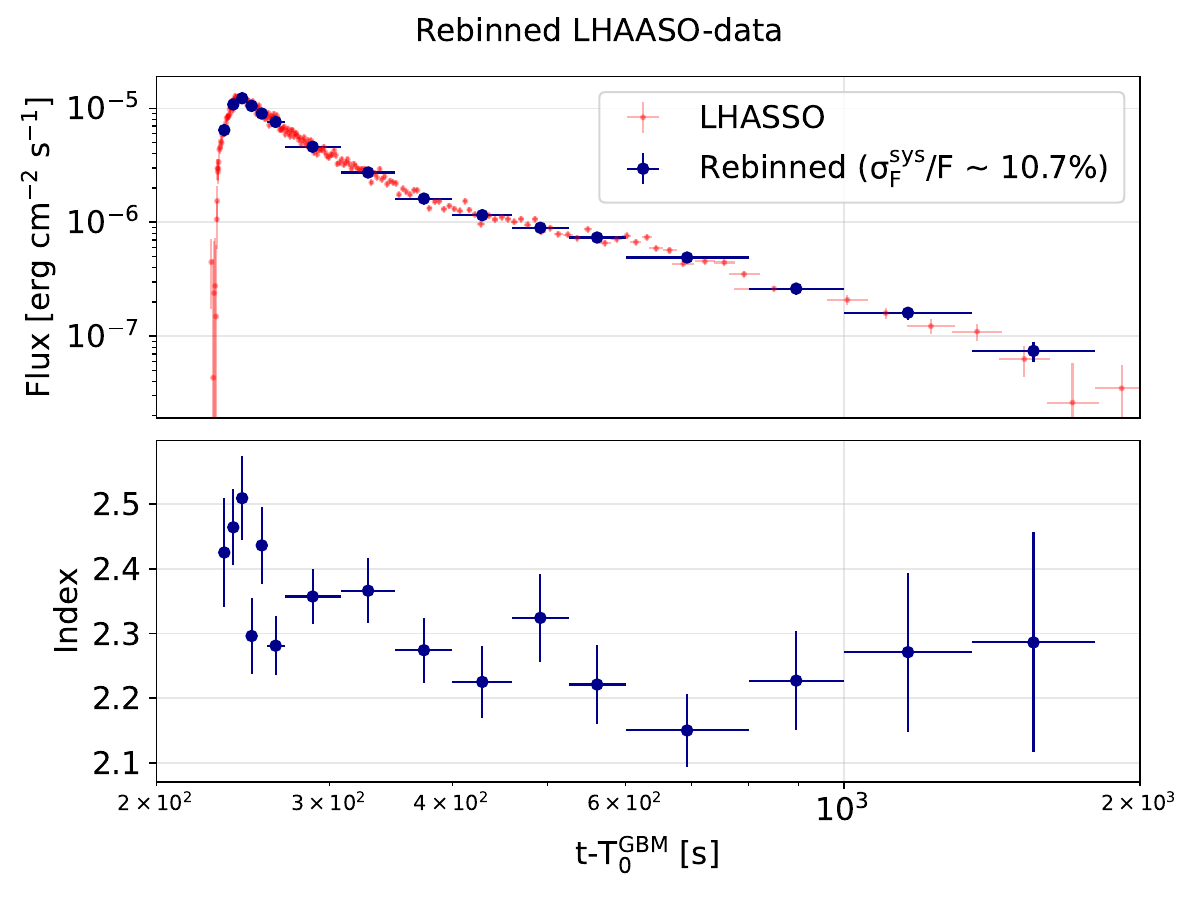}
    	\label{fig:LHAASO_rebinned}
\end{figure*}
\bmhead{LHAASO}
GRB 221009A was observed in the LHAASO field for more than 30 minutes after the initial trigger by Fermi/GBM. The detection count surpassed 64,000 photons with energies greater than 200 GeV. The peak emission was observed around 240\,s after the trigger. The observed VHE gamma-ray emission is identified as an afterglow. Spectral data, as shown in Table \ref{tab:LHAASO_rebinned_table}, was adapted from \cite{LHAASO2023}. These LHAASO bins are used as the standard for subsequent analyses adding simultaneous (or quasi-simultaneous)  MWL data from Fermi/LAT or AGILE. The information presented in Tab. \ref{tab:LHAASO_rebinned_table} includes both the flux and the intrinsic spectral index of the source, corrected for extragalactic background suppression \cite{LHAASO2023}. The intrinsic spectral index for VHE gamma-rays in the energy band of 0.3-5\,TeV is between 2.2 and 2.4, while the flux varies by two orders of magnitude (10$^{-5}$--10$^{-7}$ erg\,cm$^{-2}$\,s$^{-1}$). Although the maximum energy photon from GRB 221009A is reported to be about 10\,TeV \cite{thelhaasocollaborationVeryHighenergyGammaray2023}, energies above 5 TeV are not considered, as they do not affect the assertions made in this study. 
The (finely binned) gamma-ray flux of VHE (Fig. 3A\footnote{\url{https://www.nhepsdc.cn/files/20230518/Figure3A_4.txt}}) is obtained from LHAASO as detailed in \cite{LHAASO2023}, together with the intrinsic spectral indices (course-binned; Fig. 3B\footnote{\url{https://www.nhepsdc.cn/files/20230518/Figure3B.txt}}). These reported flux and spectral indices in Fig. 3 from \cite{LHAASO2023} have been corrected for extragalactic background light (EBL) and represent the intrinsic parameters of the source. The fluxes in the broad temporal bins, the same bins for which the spectral indices are reported, are not documented in \cite{LHAASO2023}. Therefore, we rebin the flux (utilizing the data from Fig. 3A$^{1}$) and apply a zeroth-order polynomial fit. As noted in \cite{LHAASO2023}, the published fluxes exclude the systematic uncertainty of 10.7\%. In our rebinning analysis, we incorporate this 10.7\% systematic uncertainty together with the statistical uncertainty of the fit, in quadrature. The results of this rebinning are reported in Fig. \ref{fig:LHAASO_rebinned} and Tab. \ref{tab:LHAASO_rebinned_table}.

\bmhead{Fermi/LAT}\label{sec:LAT}
We use {\sc gtburst} software from the official \textit{Fermi}-tools to extract and analyze the GRB 221009A data. The high-energy data are extracted in the energy band of 0.1-100\,GeV from the region of interest (ROI) of 12$^{\circ}$ around the source location of R.A. $= 288.28^{\circ}$ and Dec. $= 19.49^{\circ}$ \cite{GRBcoordinate}. The trigger by \textit{Fermi}/GBM on 2022-10-09 13:16:59.988 UTC (equivalent to 687014225\,s MET) is considered the trigger time (T$_{0}$). Considering that the source was bright, we have not considered any source off-axis angle ($\theta$) cut to clean the data. In addition, a zenith angle cut of 100$^{\circ}$ was applied to reduce the contamination of the gamma-ray photons from the Earth limb. 
A spectral model of type "powerlaw2" is considered for the analysis throughout along with "isotr\_template" and "template (fixed norm.)" for the particle background and the Galactic component, respectively. An "unbinned likelihood analysis" is considered with a minimum test statistic (TS$_{\rm min}$) of 10. The selection of the instrument response function is made based on the duration. We used the response function "P8R3\_TRANSIENT010E\_V2" until 400\,s from the trigger time following the analysis methods described in \cite{2019ApJ...878...52A, Bissaldi:2023gi}. 
No LAT LLE data are publicly available for this burst. We generate LAT data products through {\sc gtburst} using the Standard ScienceTool\footnote{\url{ https://fermi.gsfc.nasa.gov/ssc/}} and pipeline {\tt gtbin}. In addition, we produce background counts and response files using {\tt gtbkg} and {\tt gtrspgen}, respectively (for example \cite{ajello2020}). A cut in the angle of the off-axis source of 90$^{\circ}$ has been considered to produce the spectral files in six energy bins between 100 MeV and 100 GeV. We fit the LAT spectrum on {\sc XSPEC} using Cash statistics.  
Following the recommendation of the \textit{Fermi/}LAT collaboration \cite{Bissaldi:2023gi}, we have considered an energy threshold of 125\,MeV for BIN-2 (244\footnote{The bin mentioned in Tab. \ref{tab:MWLdata} starts from 242.1\,s, thus it covers a fraction of the entire bin of LHAASO} -247\,s) and BIN-3 (247-253\,s). The spectral data points obtained for each time-bin are consistent with the results from the standard likelihood analysis.

\bmhead{AGILE}\label{sec:AGILE}
Based on the operational status of the detectors (MCAL, RM, and GRID) on board AGILE, as detailed in Table 1 in \cite{Tavani_2023}, we have identified specific temporal bins that align with selected temporal bins presented in Table \ref{tab:MWLdata}: bin-c1: 273.0--303.0\,s; bin-c2: 303.0--383.0\,s; bin-d: 684.0--834.0\,s; bin-e: 1129.0--1279.0\,s. These bins correspond to our chosen bins: BIN-6; BIN-7; BIN-8; BIN-12; and BIN-14 (see Fig. \ref{fig:main_LC1} and Table \ref{tab:MWLdata} for details). For BIN-7 and BIN-8, the LAT data was utilized, since the AGILE data for these bins covered a longer period of 303.0--383.0\,s (bin-c2), and hence the spectra information is not publicly available for shorter exposure. 

\begin{table*}[ht]
    \centering
    \caption{The parameters of the spectral model of GRB 221009A observed with Fermi/GBM for bins which are relevant for the multi-wavelength analysis. The model parameters are presented along with the simultaneous data in Fig. \ref{fig:SED_Fig2}. Additionally, the last two columns report the peak energy and intergrated flux between 40 keV and 40 MeV. The fluxes are reported in Fig. \ref{fig:GeVandTeV} relative to the light curves of other bands.}
    \label{tab:data}
    \begin{tabular}{c c c r c c|}\hline
    \vspace{0.1cm}
             t-T$^{\rm GBM}_{0}$ [s]       & $\alpha$                 & $\beta$                 &  E$_{\rm peak}$ [keV] & Flux [erg cm$^{-2}$ s$^{-1}$] \\ \hline
             \vspace{0.15cm}
             242.00-247.00 & -1.26 $^{+0.02}_{-0.03}$ & -2.75$^{+0.04}_{-0.03}$ & 1072$^{+18}_{-18}$   & 32.70$^{+0.25}_{-0.24} \times 10^{-5}$ \\ 
             \vspace{0.15cm}
             277.00-308.00 & -1.60 $^{+0.01}_{-0.01}$ & -2.56$^{+0.09}_{-0.06}$ & 1030$^{+32}_{-36}$   & 57.20$^{+1.10}_{-0.76} \times 10^{-6}$ \\ 
             \vspace{0.1cm}
             308.02-350.28 & -1.54 $^{+0.06}_{-0.07}$ & -2.18$^{+0.01}_{-0.01}$ & 90$^{+2}_{-2}$       & 77.50$^{+1.33}_{-1.45} \times 10^{-7}$ \\ 
             \vspace{0.1cm}
             350.28-401.30 & -1.82 $^{+0.01}_{-0.01}$ & -2.47$^{+0.11}_{-0.09}$ & 312$^{+17}_{-20}$    & 38.20$^{+1.24}_{-1.32} \times 10^{-7}$ \\ 
             \vspace{0.1cm}
             600.00-800.00 & -1.73 $^{+0.05}_{-0.05}$ & -2.31$^{+0.02}_{-0.02}$ & 71$^{+3}_{-2}$       & 44.10$^{+1.20}_{-1.13} \times 10^{-8}$ \\ 
             \vspace{0.1cm}
             1000.00-1350.00 & $<$-0.16 & -2.14$^{+0.02}_{-0.02}$ & $<$53 & 19.60$^{+0.90}_{-8.75} \times 10^{-9}$ \\ \hline
        \end{tabular}
        \label{tab:GBM_SEDmodel}
    \end{table*}

\bmhead{Fermi/GBM}

Given the unprecedented duration of this burst, the response matrices directly provided by the Fermi GBM Burst catalogue\footnote{\url{https://heasarc.gsfc.nasa.gov/W3Browse/fermi/fermigbrst.html}} do not cover all the emission, allowing for an analysis up to $\sim 460$ s from the burst trigger.\\
In order to analyze time-bins at later times, we generate response matrices using the GBM Response Generator \footnote{\url{https://fermi.gsfc.nasa.gov/ssc/data/analysis/gbm/DOCUMENTATION.html}}. This official Fermi tool allows for the creation of GBM response matrices out of GBM daily data for a burst at a given location and during an arbitrary time. We download the daily data associated with October 9, 2022 from the Fermi GBM Daily Data online repository \footnote{\url{https://heasarc.gsfc.nasa.gov/W3Browse/fermi/fermigdays.html}}.\\
We produced response files for the NaI 4, NaI 8, BGO 0 and BGO 1 detectors for a source at the BOAT location (RA = $288.26^{\circ}$ and DEC = $19.77^{\circ}$, as provided by \textit{Swift}/XRT observations) between 400 s and 1800 s after GBM trigger. 
For each time-bin before 400 s, we produce weighted response files from the response matrices provided by the Fermi GBM catalog, whereas we use the custom-made response matrices for any time-bin after 400 s.
Once the GBM data are reduced, we perform spectral fits on all the selected time bins. For the fitting process, we use \textsc{pyXSPEC} \footnote{\url{https://heasarc.gsfc.nasa.gov/xanadu/xspec/python/html/index.html}}, the Python interface to the \textsc{XSPEC} spectral fitting program. We implemented the Python-based Bayesian X-ray Analysis (BXA) \cite{Buchner2016}, which allows Bayesian parameter estimation and model comparison through nested sampling algorithms in \textsc{pyXSPEC}. 
We ignored the energy channels outside 40-900 keV for the NaI detectors. We selected the energy range 400 keV - 40 MeV for the BGO data. Note that we ignored data of BGO for the BIN-14 since it is consistent with the estimated background spectrum (see below). The median values of the parameters of the Band function together with the 1-$\sigma$ intervals in their posterior distribution are reported in the Table \ref{tab:GBM_SEDmodel}.  

\begin{figure*}[ht]\caption{Comparison of the backgrounds above 1 MeV estimated with the OSV tool and the polynomial fit for GRB 221009A. 
The left panel represents the light curve above 1 MeV.  
The OSV background uses Poisson uncertainty for each bin, which is omitted graphically for visual clarity.}
\centering 
    \includegraphics[width=0.47\textwidth]{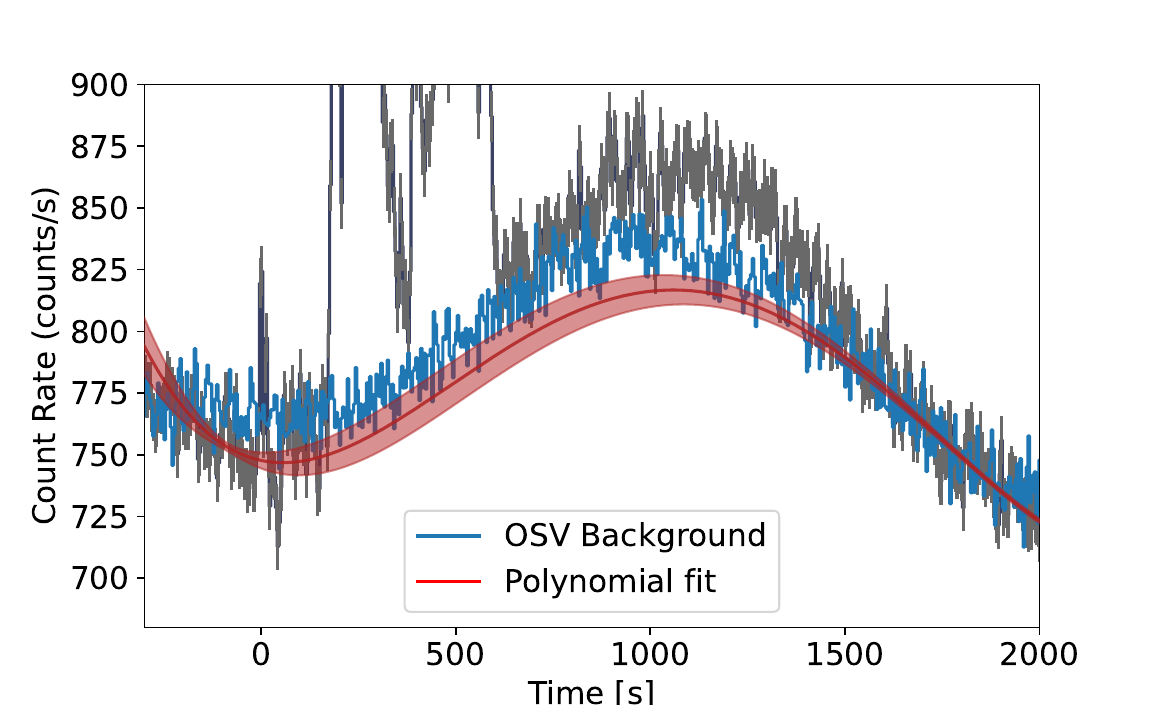}
    \includegraphics[width=0.47\textwidth]{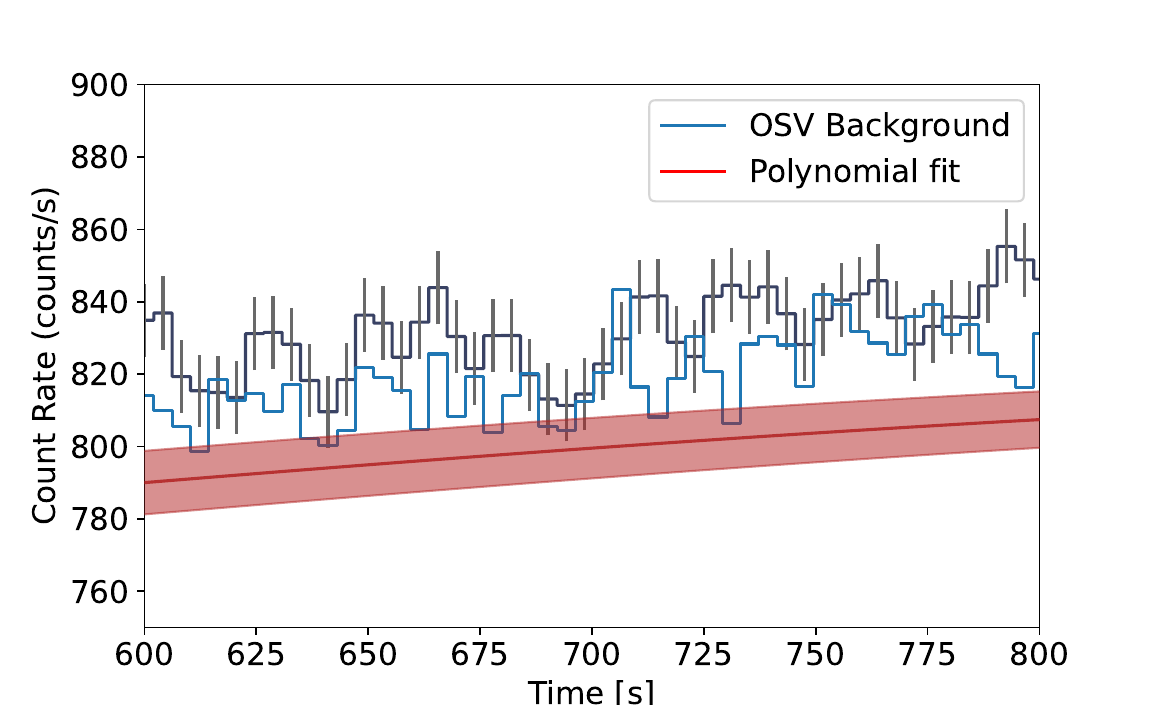}\label{appendix:fig:1MevCut}
\end{figure*}

\begin{figure}\caption{Light curve of GRB 221009A above 1 MeV with polynomial fit considering the alternative methods described in Tab.~\ref{appendix:tab:new_intervals}.}
    \centering
    \includegraphics[width=0.8\textwidth]{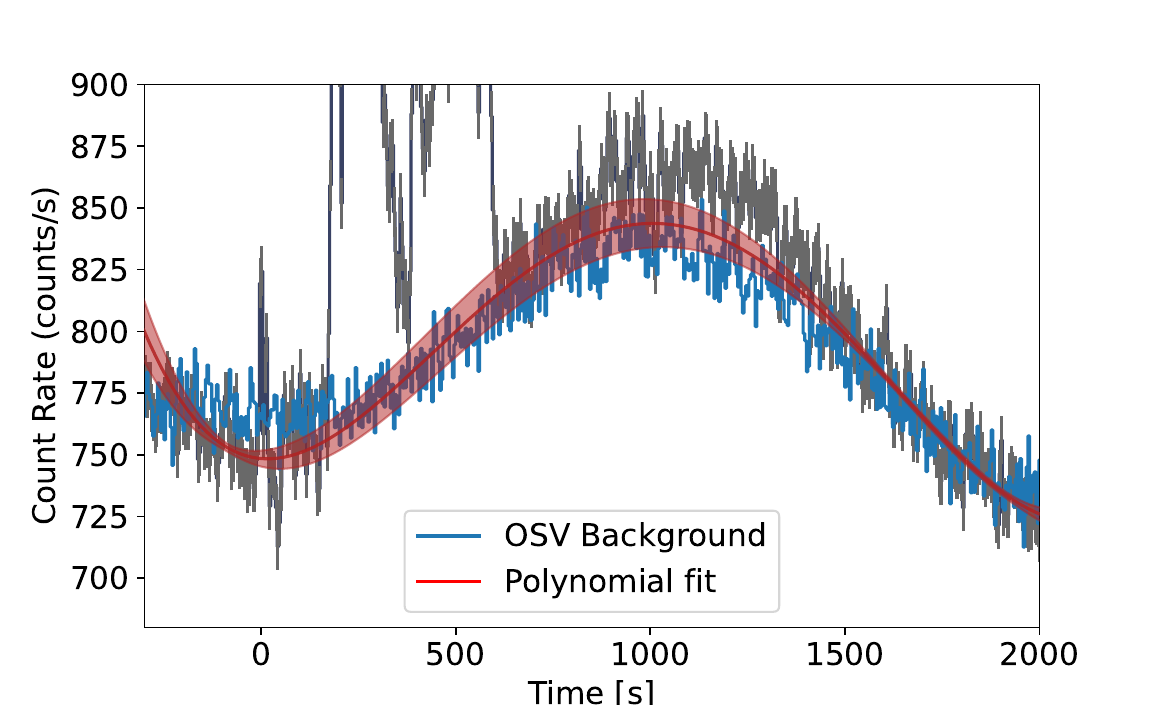}
    \label{appendix:fig:osv_fit_2}
\end{figure}

\begin{figure}[t!]\caption{The variation of peak energy as a function of time starting from 240\,s after the trigger time of GBM until about 1500\,s. No apparent variation of the peak has been observed. The red band represents the 1-sigma confidence interval of the best-fit linear (pol-0) model.}
\centering 
    \includegraphics[width=\linewidth]{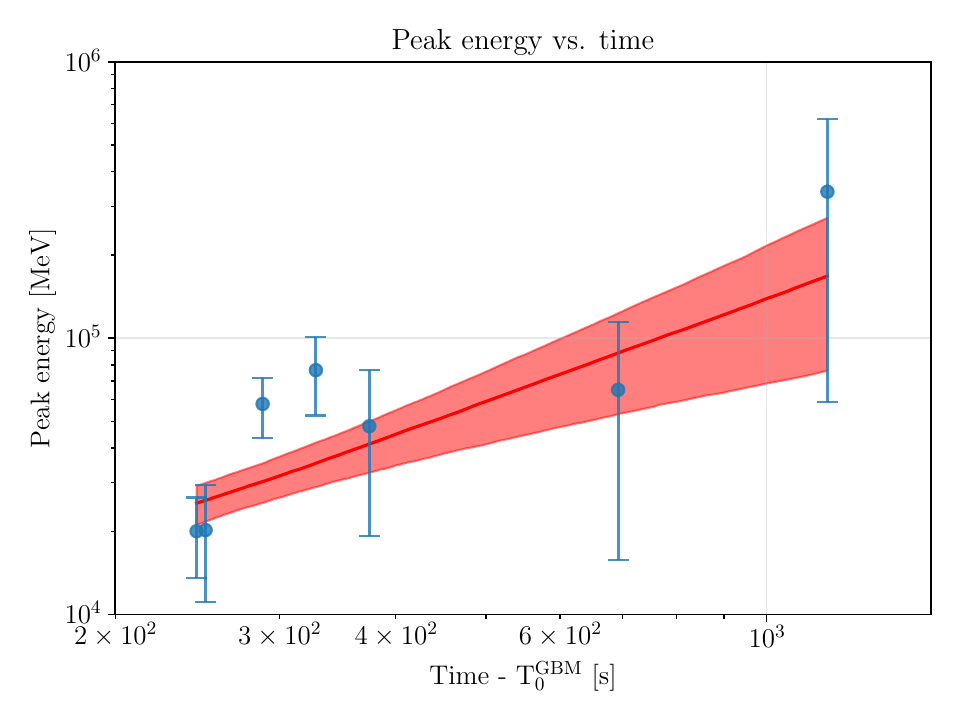}     
	\label{fig:LogParModel2}
\end{figure}

\begin{figure*}[ht]\caption{The Bolometric flux (between 10 MeV--10 TeV) of the second component estimated for time bins mentioned in Table \ref{tab:goodbins}. }
\centering 
    \includegraphics[width=\linewidth]{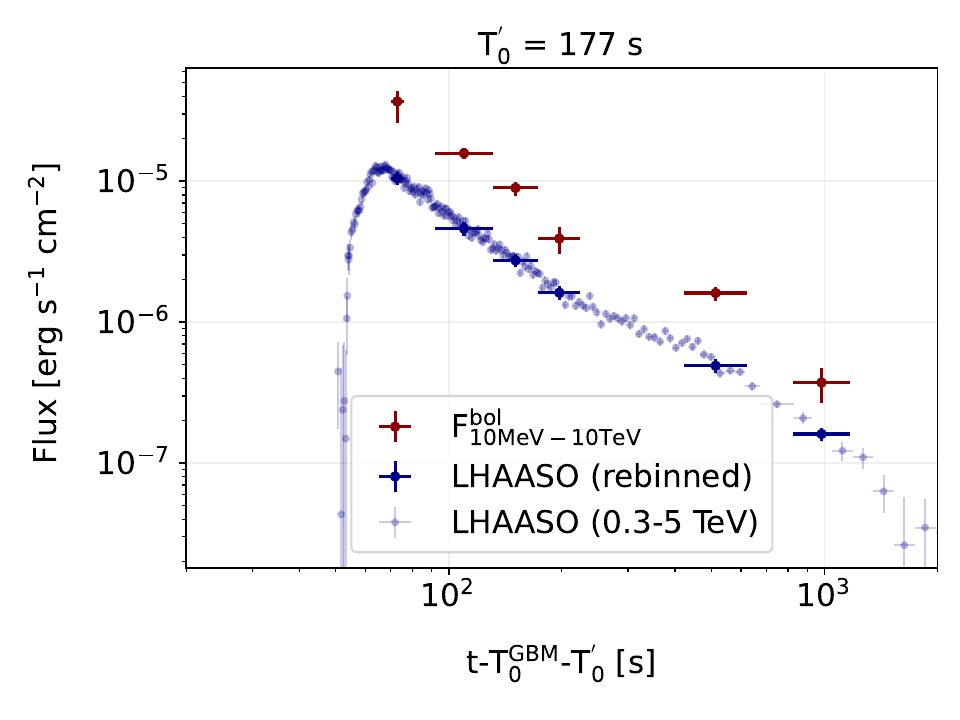}
    	\label{fig:VHE_FBol}
\end{figure*}

\bmhead{GBM background}
Given that the MeV emission of GRB 221009A lasts more than 1000\,s, using a lower-order polynomial (polynomial of order 2-3) to model the MeV background is ineffective. This ineffectiveness is due to the significant variations in the MeV background throughout the GRB duration. Furthermore, the components that influence the background exposed to the detector exhibit considerable fluctuations over these periods \cite{phys_model_gbm}. Accurate modeling of the background for such extended duration requires the use of higher-order polynomials, which, however, raises the risk of data overfitting. A different method for estimating the background was originally introduced by \cite{fitzpatrick_osv} via the Fermi Orbital Subtraction Tool (OSV). 
The GBM background radiation exposure varies in terms of position and orientation, as the satellite oscillates every two orbits covering full-sky over 30 orbits. By calculating the average rates from 30 orbits prior and subsequent, we derive a background estimate for the desired time. Using the background intervals specified in \cite{Lesage_2023} for a polynomial fit, we evaluated these methodologies. As described in Fig.~\ref{appendix:fig:1MevCut}, left panel, the polynomial fit tends to underestimate the background compared to the estimation of the OSV tool. Consequently, in the range of 600-800~s (BIN-12 from Tab.~\ref{tab:LHAASO_rebinned_table}), traditional analysis (prescribed in \cite{Lesage_2023}) tends to overestimate the MeV signal due to inaccurate background calculations. In this range, the polynomial fit gives around $\sim$6000 excess counts while the OSV tool estimates an excess of $\sim$2000 (see Fig.~\ref{appendix:fig:1MevCut}, right panel). The sensitivity of the polynomial fit to the chosen background region likely explains this discrepancy. Table~\ref{appendix:tab:new_intervals} shows the (approximate) intervals chosen by \cite{Lesage_2023} along with a different set of background ranges used for the analysis performed in this work. The latter considers that post-1470\,s, the source is occulted by Earth, enabling a selection of ranges closer to the true signal. This polynomial fit predicts a higher background than the OSV tool, as shown in Fig.~\ref{appendix:fig:osv_fit_2}. Since the OSV tool exhibits less sensitivity to background region selection compared to polynomial fitting, it presents a reliable alternative for estimating backgrounds in long-duration ($\sim ks$) events. 
The development of the OSV pipeline used in this work utilizes the foundational tools outlined in \cite{fitzpatrick_osv}. In order to adapt to the requirements, certain components of the code have been modified. The primary changes involve migrating from Python 2 to Python 3. The complete code is available at the following link: \cite{De_Santis_2024}.

\begin{table}[ht]\caption{Two different possible selections for the background regions. Taking into account that the Earth occludes the source after 1470s.}
        \centering 
        \begin{tabular}{ c c c}\hline
          Cut &  Left [s] & Right [s] \\  \hline
          Lesage et al. 2023 \cite{Lesage_2023} & (-150, -10) & (1625, 2400) \\
          Alternative (This work) & (-150, -10) & (1500, 2000) \\
          \hline
        \end{tabular}
        \label{appendix:tab:new_intervals}
\end{table}

\bmhead{Konus-\emph{Wind}}
The Konus-$Wind$ (KW) detection of GRB~221009A and the data reduction procedures are described in detail in \cite{Frederiks_2023}. In this work, we used dead-time and saturation-corrected KW spectral data (20~keV -- 20~MeV) for five consecutive time intervals: 216.832-225.024, 225.024-233.216, 233.216-241.408, 241.408-249.600, 249.600-257.792~s after the trigger (i.e., spectra \# 60-64, as listed in Table~1 in \cite{Frederiks_2023}). The median values of the parameters of the Band function together with the 1-$\sigma$ intervals in their posterior distribution are reported in Table \ref{tab:GBM_SEDmodel}.  

\bmhead{Selection of relevant temporal bins in HE and VHE gamma-rays}
Selection of appropriate temporal bins is carried out for the joint HE and VHE gamma-ray analysis to ensure the coverage of the energy band ranging from 100 MeV to 5 TeV. Of all available bins with LHAASO spectral data, only seven bins meet the criteria for high-quality MeV-TeV analysis. The rebinnned spectral data from LHAASO (see Methods for details) is documented in Table~\ref{tab:goodbins}. The Fermi/LAT spectral information is processed following the method described in the LAT section in Methods. For combined HE and VHE spectral analysis, the following bins: BIN-2, -3, -6, -7, and -8 have been chosen with Fermi/LAT and LHAASO datasets. For bins, BIN-12 and BIN-14, data from AGILE, obtained from \cite{Tavani_2023} (see the AGILE section in Methods for details), along with LHAASO.
\begin{equation}
    \rm \frac{dN}{dE} = N_0 \left(\frac{E}{E_0} \right) ^{-\alpha - \beta log (E/E_0)} = N_{\rm p} \exp \left( - \frac{\log^2(E/E_{\rm p})}{w} \right)
\end{equation}\label{eq:LogPareq}

\bmhead{Empirical modeling of the second component}\label{appendix:modelLP}
A second spectral component consisting of high-energy gamma rays (GeV) and very-high-energy gamma-rays (TeV) is usually characterized by a Log-Parabolic (LP) spectral shape for jetted objects. In the past, this method has been applied for several VHE objects, such as Crab Nebula \cite{Aleksi__2015}; Mrk501 \cite{Massaro_2006}; bright VHE blazars detected with HAWC \cite{2023JCAP...10..009J}; and Fermi/LAT blazars with VHE emission \cite{Zhou_2021}. Compared to the power-law model (defined as $dN/dE\, \propto (E/E_0)^{ - \alpha}$ one additional parameter is considered ($\beta$) which describes the curvature of the spectrum. The functional form is defined as follows: \newline
This is a simple phenomenological model showing a peak at an energy $E_{\rm p} = E_0 \exp(- \alpha / 2 \beta)$. 
In order for the posterior distributions in our analysis to be more intuitively meaningful, we re-parameterize the model with the peak energy $E_{\rm p}$, the width $w = 1/\beta$, and the peak flux $N_{\rm p} = N_0 e^{\alpha^2 / 4 \beta}$.
Since the aim is to probe the intrinsic parameter space using an empirical methodology, intrinsic spectral parameters are considered for the VHE gamma-ray spectrum from LHAASO which are corrected for EBL (mentioned in Table \ref{tab:LHAASO_rebinned_table}). The high-energy gamma-ray data have been obtained from Fermi/LAT and AGILE, whenever available. GRB 221009A moved out from the Fermi/LAT FoV after about 400\,s. However, thanks to AGILE, after 400\,s, the GeV band is still covered in the energy bin of 50 MeV to 50 GeV. The analysis was performed in the energy bin of more than 50\,MeV-5\,TeV for the joint analysis between GeV (AGILE/ LAT) and LHAASO.\\
The LHAASO spectral data for different temporal bins are not publicly available. However, 
in the energy bin of 300 GeV to 5 TeV the intrinsic spectral parameters, such as power-law index and intrinsic energy flux, are available from \cite{LHAASO2023}. Assuming that the distribution of these two parameters is Gaussian, a posterior predictive distribution (PPD) has been estimated (see Fig.~\ref{fig:SED_Fig2}).
We developed a technique to extract synthetic data points from the PPD defined by the posterior distribution on the spectral parameters of the data from LHAASO. We validated this technique by performing a power law fit of the synthetic data points, which recovers the original posterior distribution on the fit parameters.
The procedure is as follows: A power law model is defined as ${F(E; a)}$, where ${a}$ denotes the model parameters (say, the bolometric flux over the energy band and the spectral index). 
The starting information is a distribution ${p(a)}$; using this, for each energy ${E}$ we calculate a PPD:
$F(E; a)\vert_{a \sim p(a)}$,
approximate this distribution as a Gaussian, and estimate the mean ${\overline{F}(E)}$ and standard deviation ${\sigma(E)}$ from the posterior samples. Finally, we generate four geometrically spaced synthetic data points in the energy band of 0.3-5\,TeV. For each of the points, we assign a flux $\overline{F}(E)$ and a standard deviation $\sigma(E) \sqrt{n/d}$, where $d=2$ is the dimensionality of our model while $n=4$ is the number of synthetic data points. 

Deriving the "synthetic spectral data" based on the spectral parameters alone will lead to a loss of the information because it will erase any evidence of the spectral curvature in the LHAASO band alone. However, after correcting for EBL absorption, the spectrum is compatible with a power-law \cite[Fig.\ 3]{thelhaasocollaborationVeryHighenergyGammaray2023} in the energy range 0.3-5\,TeV: the loss of information is minimal.
The evidence for curvature we then find arises entirely from considering the TeV data together with GeV data.

\bmhead{Combined light curve in GeV energies with AGILE and Fermi/LAT}
LHASSO has identified a break around 600\,s (from the GBM trigger time) in the light curve of GRB 221009A within the energy range of 0.3-5 TeV. The apparent featureless light curve seen in VHE gamma rays is considered indicative of a jet break. To investigate the cause of the slope change at 600\,s post-GBM trigger, we generated a light curve for HE gamma-rays using data from both LAT and AGILE. We selected an energy range from 125 MeV to 3 GeV, which is covered by both instruments. This choice of 125 MeV adheres to the Fermi/LAT collaboration recommendation as outlined in \cite{Bissaldi:2023gi}. Flux measurements in the 0.125-3 GeV range are presented in Table \ref{tab:LAT_eneLC}. The data from AGILE that were initially in photon flux are converted to energy flux as detailed in Table \ref{tab:AGILE_eneLC}, assuming photon indices of $-2.0$\footnote{This assumption of a photon index of $-2.0$ is consistent with \cite{Tavani_2023} to analyze GRID observations in the energy range of 50 MeV to 3 GeV as described in Section B.1}. 
The conversion from the photon flux to the energy flux is performed using equations mentioned below:
\begin{equation}
    \rm F^{[E_1, E_2]}_{E} = F_{ph} \times log\left( \frac{E_2}{E_1}\right) \times  \frac{E_1 E_2}{E_2-E_1}, 
\end{equation}\label{eq:AGILE_e_corr1}

where F$\rm ^{{E_1, E_2}}\rm_{E}$ is the energy flux between the energy range of {E$_1$, E$_2$}. The energy flux in the band of [125 MeV, 3 GeV] is obtained by the following scaling factor:\newline

\begin{equation}
 {\rm F}^{ \rm 0.125-3\,GeV}_{\rm E} = \rm log\left(\frac{30}{0.125}\right) \times \frac{\rm F^{E_1, E_2}_{E}}{ \rm log(E_2/E_1) }  \newline 
 = 5.48 \times  \rm F^{[E_1, E_2]}_{ph}  \times  \frac{E_1 E_2}{E_2-E_1}
\end{equation}\label{eq:AGILE_e_corr2}

The light curve is presented starting from a trigger time of 177\,s, marking the detection of the initial gamma-ray pulse by KONUS \cite{Frederiks_2023}. \\
Furthermore, we have incorporated low-energy data from Fermi/GBM in the 40 keV-40 MeV energy range (see  Fig. \ref{fig:GeVandTeV}). The keV-MeV emissions are classified into two time segments: the prompt phase (duration 100\,s) and the combined prompt and afterglow phase (100-1000\,s).
It is observed that the combined prompt and afterglow emission phase post 600\,s aligns with the temporal decline observed in the HE/VHE gamma-rays.

\begin{table*}[ht]\caption{The energy flux of AGILE. The photon flux as observed by AGILE has been converted to the energy flux as prescribed in Eq. \ref{eq:AGILE_e_corr1} and \ref{eq:AGILE_e_corr2}. The photon fluxes for this purpose have been obtained from Table 3, 4, and 5 in \cite{Tavani_2023}. The energy-flux is further used in the Fig. \ref{fig:GeVandTeV}.}
        \centering 
        \begin{tabular}{ c c c c c c }\hline
        \multicolumn{4}{c}{AGILE }                                                            \\ \hline
           \multirow{2}{*}{Time - T$_0^{\rm GBM}$ [s]}   &  Energy range &F$_{\rm ph}$& ${\rm F}^{ \rm 125\,MeV-3\,GeV}_{\rm E}$ \\ 
                      &  [MeV]         & [ph\,cm$^{-2}$\,s$^{-1}$] & [erg\,cm$^{-2}$\,s$^{-1}$] \\ \hline
           273.0-303.0 & 50–3000       & (1.5$\pm$0.2)10$^{-2}$   &(3.88$\pm$0.52)10$^{-6}$      \\
           303.0-383.3 & 50–3000       & (5.4$\pm$0.6)10$^{-3}$   &(1.40$\pm$0.16)10$^{-6}$      \\ 
           303.0-383.3 & 50–3000       & (5.4$\pm$0.6)10$^{-3}$   &(1.40$\pm$0.16)10$^{-6}$       \\  
           684.0-834.0 & 50–3000       & (9.0$\pm$1.7)10$^{-4}$   &(2.85$\pm$0.52)10$^{-7}$      \\ 
           1129.0-1279.0  & 50–50000       & (1.7$\pm$0.8)10$^{-4}$   & (4.33$\pm$2.04)10$^{-8}$     \\ 
           1569.0-1719.0  & 50–50000       & (1.0$\pm$0.5)10$^{-4}$   & (2.55$\pm$1.27)10$^{-8}$      \\ 
           2014.0-5269.0  & 50–3000        & (9.5$\pm$5.2)10$^{-5}$   & (2.46$\pm$1.35)10$^{-8}$\\ 
           5273.0-16980.0 & 50–3000        & (2.6$\pm$1.2)10$^{-5}$   & (6.73$\pm$3.11)10$^{-9}$\\ \hline
        \end{tabular}
        \label{tab:AGILE_eneLC}
    \end{table*}

\begin{table*}[ht] \caption{The observed flux in the 0.1-3 GeV energy band observed by Fermi/LAT. The selection of this energy range is intended to construct a combined GeV light curve covering from 240\,s to 40\,ks.}
        \centering 
        \begin{tabular}{ c c }\hline
        \multicolumn{2}{ c }{LAT (125\,MeV-3\,GeV)} \\ \hline
        Time-T$_0^{\rm GBM}$ [s]                     & Flux [erg\,cm$^{-2}$\,s$^{-1}$] \\  \hline
244.0--247.0 &(3.25$\pm$0.61) 10$^{-5}$\\ 
247.0--253.0 &(2.88 $\pm$ 0.43) 10$^{-5}$\\ 
277.0--308.0 &(4.59 $\pm$ 0.60) 10$^{-6}$\\ 
308.0--350.0 &(2.66$\pm$ 0.43) 10$^{-6}$\\ 
350.0--401.0 &(1.49 $\pm$ 0.35) 10$^{-6}$\\ 
3893.0--6143.0 &(5.95 $\pm$ 0.95) 10$^{-9}$\\ 
9593.0--11843.0	&(2.32 $\pm$ 1.06) 10$^{-9}$\\ 
15294.0--17544.0 & (1.13$\pm$ 0.48) 10$^{-9}$\\ 
20000.0--40000.0 & (8.06 $\pm$ 2.00) 10$^{-10}$ \\ \hline
\end{tabular}
       \label{tab:LAT_eneLC}
    \end{table*}

\begin{table*}[ht!]
\caption{Coverage of GRB 221009A by MWL instruments, such as LHAASO, Fermi/GBM, -LAT, KONUS and AGILE. The temporal bins are driven by the observation of LHAASO \cite{LHAASO2023}, and are numerically named from BIN-0 to BIN-14. The coverage (/availability) of the other instruments are marked as "\cmark" in the case of available usable data or "\xmark" in case of unavailability (or unusable data due to pile-up, for example) of the data. Fermi/GBM MeV data are only selected during the period when there is no pile-up \cite{Lesage_2023}. The energy band between 30 MeV to above 100\,GeV is covered by both AGILE and Fermi/LAT. In the case of AGILE the quasi-simultaneous time-periods presented in \cite{Tavani_2023} are selected for multiwavelength time-resolved spectral analysis.}
\begin{tabular}
        { c c c c c c c}\hline
         Bin  &  t-T$^{\rm GBM}_0$ [s]         &  LHAASO  &  GBM           & AGILE  &  LAT    &  KONUS\\ \hline
           BIN-0  &  232.0-237.0&  \cmark  & \xmark         & \xmark &  \xmark & \cmark\\  
           BIN-1  &  237.0-242.0&  \cmark  & \xmark         & \xmark &  \xmark & \cmark\\  
           BIN-2  &  242.0-247.0&  \cmark  & \xmark         & \xmark &  \cmark & \cmark \\ 
           BIN-3  &  247.0-253.0&  \cmark  & \xmark         & \xmark &  \cmark & \xmark \\ 
           BIN-4  &  253.0-259.0&  \cmark  & \xmark         & \xmark &  \xmark & \xmark\\ 
           BIN-5  &  259.0-270.0&  \cmark  & \xmark         & \xmark &  \xmark & \xmark\\ 
           BIN-6  &  270.0-308.0&  \cmark  & \cmark         & \cmark / bin-c1  &  \cmark& \xmark \\ \hline
           BIN-7  &  308.0-350.0&  \cmark  & \cmark         & \multirow{2}{*}{\xmark/ bin-c2} & \cmark & \xmark\\
           BIN-8  &  350.0-401.0&  \cmark  & \cmark         &                           & \cmark & \xmark \\ \hline
           BIN-9  &  401.0-463.0&  \cmark  & \cmark         & \xmark & \xmark  & \xmark \\ 
          BIN-10  &  463.0-525.0&  \cmark  & \xmark & \xmark & \xmark  & \xmark \\ 
          BIN-11  &  525.0-600.0&  \cmark  & \cmark         & \xmark & \xmark  & \xmark \\ 
          BIN-12  &  600.0-800.0&  \cmark  & \cmark         & \cmark/ bin-d   & \xmark & \xmark \\ 
          BIN-13  &  800.0-1000.0&  \cmark & \cmark         & \xmark           & \xmark & \xmark \\ 
          BIN-14  &  1000.0-1350.0&  \cmark& \cmark         & \cmark/ bin-e   & \xmark & \xmark \\ \hline
        \end{tabular} \label{tab:MWLdata}
    \end{table*}

\begin{table*}[ht!]
\caption{Time coverage of different instruments gathered to perform the time-resolved spectra of GRB 221009A. The designated reference time bins (labeled BIN-0 to -15, see Table \ref{tab:MWLdata}) correspond to the LHAASO bins, chosen for their uniform data availability that extends nearly an hour after the GBM trigger time (T$^{\rm GBM}_0$). The selection of MWL data from various instruments depends on their coverage in these reference time bins. Due to the high fluence of GRB 221009A, the GBM and LAT detectors suffered from pile-up issues, resulting in incomplete coverage of the LHAASO bins. Additionally, publicly accessible data partially include BIN-12 and -14. The bins marked with ($^*$) in the first column denote partial coverage.} \centering 
        \begin{tabular}{lccccc}\hline
            BIN   &   LHAASO [s]   &  LAT [s]    &  AGILE [s]   &   GBM [s]      & KONUS [s]    \\ \hline
           2$^*$  &  242.0-247.0   & 244.0-247.0 &\xmark        & \xmark         & 241.4-249.6 \\ 
           6$^*$  &  269.6-308.0   & 277.0-308.0 & \xmark       & 277.0-308.0    & \xmark     \\ 
           7      &  308.0-350.0   & 308.0-350.0 &\xmark        & 308.0-350.0    & \xmark  \\  
           8      &  350.0-401.0   & 350.0-401.0 &\xmark        & 350.0-401.0    & \xmark   \\ 
          12$^*$  &  600.0-800.0   & \xmark      &684.0-834.0   & 600.0-800.0   &  \xmark       \\    
          14$^*$  &  1000.0-1350.0 & \xmark      &1129.0-1279.0 & 1000.0-1350.0 &  \xmark            \\ \hline
        \end{tabular}     
        \label{tab:goodbins}
    \end{table*}

\bmhead{Extrapolation of prompt emission component in the GeV band}
Data from the temporal bins BIN-2, -6, -7, -8, -12, and -14 are used to construct multiwavelength spectra in the range from keV to TeV. As shown in Fig. \ref{fig:GeVandTeV} and Table \ref{tab:MWLdata}, depending on availability, the GBM and KONUS data are chosen that cover the energy range of several tens of keV to several tens of MeV. Specifically for BIN-2, KONUS data is used, since GBM data was compromised by pile-up issues during this interval. For the remaining bins, we used the GBM data. The methodology for data reduction is described in Methods. Using the keV-MeV band data, we derived the optimal model parameters (see Table \ref{tab:GBM_SEDmodel}). To assess the impact of the apparent initial peak within the GeV-TeV range, the spectrum was extended to 100 GeV. 
The extended models are represented by the blue shaded areas in Figure \ref{fig:SED_Fig2}. 
The low-energy (sub-GeV) spectral data from AGILE and LAT might be influenced by the Band-function at the keV-MeV range as those overlaps with the model extended up to 100 GeV (see Fig. \ref{fig:GeVandTeV}). However, in bins such as BIN-2, -6, -7, -8, and -12, the observations above 1 GeV significantly surpass the extrapolated "first hump", suggesting the presence of an additional spectral component at GeV-TeV energies. For BIN-14, spectral data ranging from 50 MeV to 1 GeV are adequately described by a single component that extends to several tens of GeVs. The upper limit set by AGILE at approximately 20 GeV introduces a provision for a possible second component, correlating with the spectrum seen by LHAASO at even higher energies (above 300 GeV).

\bmhead{Synchrotron self-Compton model}

To model the late-epoch (1000-1350\,s) 40\,keV - 5\,TeV spectrum of the afterglow radiation of GRB 221009A, we used the Lepto-Hadronic Modeling Code \cite[LeHaMoC;][]{lehamoc}. Specifically, we adopt the leptonic module of LeHaMoC called LeMoC. We assume that the spectrum at the latest epoch is dominated by the synchrotron and inverse Compton radiation of a power-law distributed electrons accelerated in the forward shock. LeMoC allows us to estimate the energy losses of electrons/ positrons due to synchrotron radiation, inverse Compton scattering, synchrotron self-absorption, and photon-photon pair creation. The injection of particles is placed in a blob of radius $R_e$, in a magnetic field of B. The following 6 free parameters are used to model the 2 component keV-TeV spectrum of the afterglow:
the strength of the magnetic field in the comoving frame B, the minimum Lorentz factor of the injected electrons $\gamma_m$, the maximum energy of the injected electrons $\gamma_{max}$, the shape of injected electrons spectrum p ($dN/d\gamma \propto \gamma_e^{-p}$), the electron injection compactness $l_e = L_e \sigma_T/4\pi R_e m_e c^{3}$ ($L_e$ is the injection luminosity of electrons) and the Doppler factor $D$. 
We use the rough approximation for $R_e$ as $\Gamma c T_{obs}^{\prime}$ and $\Gamma \approx D/2$, where $T_{obs}^{\prime} = T_{obs}-177$ s. 

We use the \textbf{emcee} python package\cite{emcee} to sample the posterior probability density with the Markov Chain Monte Carlo approach. Information on priors and posteriors is listed in Table \ref{tab:bestfitmodel}. A corner plot with the posterior distribution of 6 parameters is shown in Fig.\ref{fig:cornerSSC1}. The posterior distibution of inferred equipartition parameters (for the homogeneous medium) is shown in Fig. \ref{fig:cornerSSC2}. 
We estimate the SSC spectra (Fig.\ref{fig:Extended_model}) and the lightcurve of the TeV emission (Fig.\ref{fig:SSC_LC}) for earlier epochs of observations by re-scaling the SSC parameters. For the homogeneous medium, beyond the deceleration time,  we expect $\Gamma \propto t^{-\frac{3}{8}}$, $R_e \propto t^{\frac{5}{8}}$, $\gamma_m \propto t^{-\frac{3}{8}}$, $B \propto t^{-\frac{3}{8}}$, $\gamma_{max} \propto t^{\frac{3}{16}}$, $l_e \propto t^{\frac{1}{2}}$. 

We also estimate the expected soft X-ray (0.5-5 keV) fluence in the afterglow phase for the first 20 minutes to be $\sim 6 \times 10^{-4} \, erg \, cm^{-2}$, which is one order of magnitude less than required to explain the observed X-ray rings from GRB 221009A \cite{Tiengo2023}. The prompt emission in the brightest pulses has flux $\sim 6 \times 10^{-6} \, erg \, cm^{-2} \, s^{-1}$ at 0.5-5 keV, once the Band function is extrapolated down to soft X-rays. Hundreds of seconds of prompt emission together with the early X-ray afterglow emission should be reasonably enough to account for the observed X-ray rings. 

\bmhead{External injection of the prompt emission photons}

In the first few hundred seconds, the MeV prompt emission is luminous (up to $\sim 10^{56}$ erg/s). In the standard prompt emission model for internal shocks \cite{Rees1994}, the MeV radiation is far behind the forward shock. Thus, the supply of prompt emission photons to electrons accelerated in the forward shock and the photons produced by them can not be ignored. There are several effects from the long-lasting ($\sim$ 500 s) energy and photon supply: (1) additional energy injection into the decelerating blast wave \cite{Derishev2024}, (2) $\gamma-\gamma$ attenuation of TeV photons \cite{Khangulyan2024}, (3) suppression of electron cooling by MeV photons and gain in photons from External Inverse Compton (EIC) scattering of X-ray photons of the prompt emission.  

There is a crucial difference between the two sources of the seed photons for the inverse Compton cooling of the forward shock-accelerated electrons and the subsequent pair production. The expected synchrotron photons from the forward shock electrons are softer, whereas the prompt are harder photons are more. Let us assume similar bulk Lorentz factors $\Gamma$ for the MeV region and the forward shock. The target photons for the pair production of observed $0.5$ TeV would be $E_t \approx 0.4 \, \Gamma_{3}^{2} E_{0.5 TeV}^{-1}$ MeV photons of the prompt emission. The optical depth for
pair production is then $\tau_{\gamma \gamma} \approx 6 \times 10^{-1} L_{\gamma,56} \Gamma_{2.7}^{-5} t_{1}^{-1} E_{\gamma, 1 MeV}^{-1}$, where $t$ is the observed time. Thus, the initial rise of the afterglow radiation could have been suppressed, which could account for the steep rise of the LHAASO light curve \cite{Khangulyan2024,Gao2024,ZhangT2023} (see also \cite{Shen2024}). It can also slightly affect TeV emission at later times, beyond the peak of the afterglow 
$\tau_{\gamma \gamma} \propto L_{\gamma} t^{\frac{7}{8}} E_{\gamma}^{-1}$ (homogeneous medium). 

On one hand, the injection of prompt emission photons suppresses the TeV afterglow radiation. On the other hand, the External Inverse Compton (EIC) of the prompt emission photons
add up photons, but with a significant delay with respect to the MeV injection time.
The EIC radiation is anisotropic and most of the photons are received at the angle $\sim 1/\Gamma$ with respect to the line of sight \cite{Aharonian1981,Brunetti2000,Beloborodov2005,Fan2006,Fan2008,2021ApJ...908L..36Z,2021ApJ...920...55Z}. Therefore, also the peak energy of EIC will be further reduced. Consequently, the observed flux of the EIC will also be suppressed due to anisotropy. We expect the EIC radiation to have a duration of $\sim R/c\Gamma^{2} \sim 4t$ (homogenous circumburst medium) with respect to the observed time of prompt emission photons, where $R$ is the size of the forward shock at a given observer time t \cite{Beloborodov2005,Fan2006}. Therefore, the upscattered prompt emission photons lose their variability in the TeV $\gamma$-rays. The effective EIC scattering is possible only at the early epochs, because the energy density of the prompt emission photons is $\propto L_{\gamma} t^{1/2}$ (homogeneous circumburst medium). 

At any time phase of the forward shock, the observed photons of the prompt emission of energy $E_{\gamma,MeV}^{obs}$ are seen as $\approx 40 E_{\gamma,MeV}^{obs} \epsilon_{e,-2}$ MeV photons in the rest frame of the typical electrons accelerated by forward shock. Therefore, only the prompt emission photons up to $E_{\gamma}^{obs} \sim 12 \epsilon_{e,-2}^{-1} $ keV will contribute to the EIC radiation. At the coasting and deceleration time of the forward shock, the emitting region of the MeV photons is expected to have comparable $\Gamma$ of the forward shock; therefore, we can compare the X-ray flux of the prompt emission with the one expected from the afterglow radiation. The prompt emission provides $\sim 10$ more photons at $\sim$ 10 keV. However, the observed delayed EIC radiation is suppressed by factor $\sim 0.5$ ($\Gamma>100$), the EIC provides slightly more GeV photons. More importantly, because of the presence of MeV photons of the prompt emission, the peak of the EIC radiation will be located at energies lower than those expected in the SSC model (less MeV photons). In other words, the external field of MeV photons significantly softens the early TeV spectra. 

\begin{figure*}[ht!]\caption{
Predicted SEDs for the early epoch at 270-308 s (BIN-6) with SSC and external injection of the prompt emission photons. The SEDs are estimated from the posterior distribution of the SSC model for the latest epoch at 1000-1350 s (BIN-14), with the self-similar dynamics of the relativistic blast wave in the homogenous (dark red) circumburst medium. A Band function with properties of the early Konus-WIND fits is chosen for the prompt emission injection into the forward shock. Note that these are not the fits to the data but predictions of the SEDs from the scaling of microphysical parameters according to the expected dynamics of the forward shock}
\label{fig:Extended_model2}
\centering 
    \includegraphics[width=0.9\linewidth]{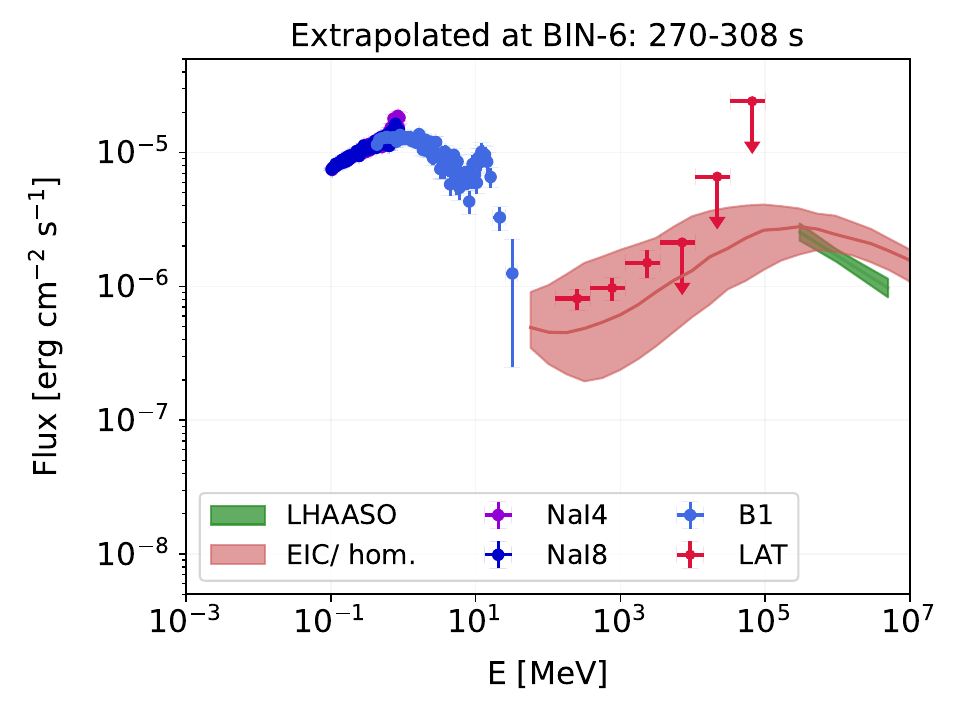}
\end{figure*}

\begin{figure}[t!]\caption{Corner plot showing the one-, two- and three-sigma contours of the joint two-dimensional posteriors for each parameter pair for SSC model of the keV-TeV spectrum at 1000-1350 s. }
\centering 
    \includegraphics[width=\linewidth]{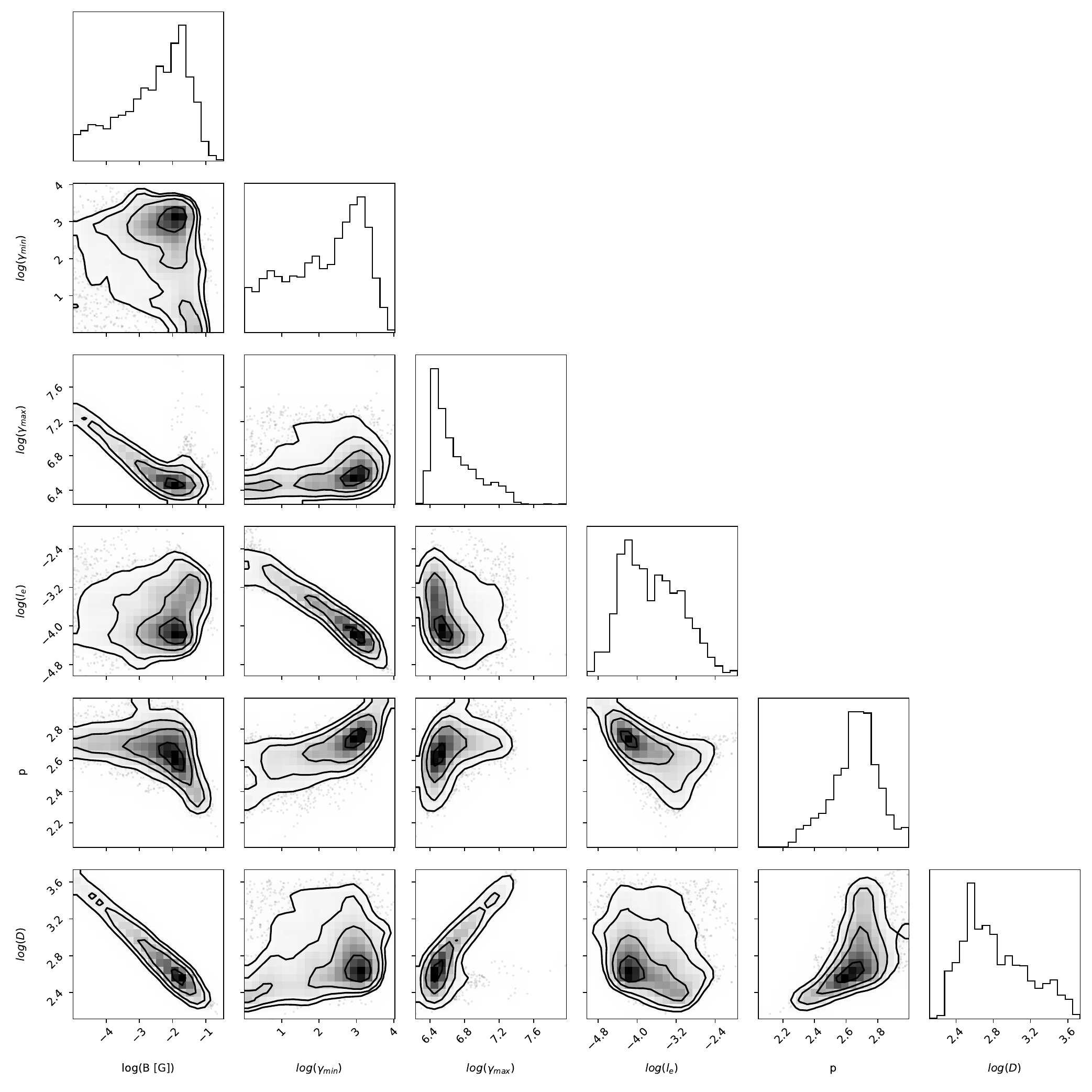}     
	\label{fig:cornerSSC1}
\end{figure}

\begin{figure}[t!]\caption{Corner plot showing the one-, two- and three-sigma contours of the equipartition parameters of the forward shock in the homogenous medium. They are inferred from the SSC modelling of the keV-TeV spectrum at 1000-1350 s. Note that the values of $\epsilon_B$ depend on the circumburst medium density $\epsilon_B \propto n^{-1}$. The posteriors of $\epsilon_B$ are here fixed to $0.1 cm^{-3}$. }
\centering 
    \includegraphics[width=\linewidth]{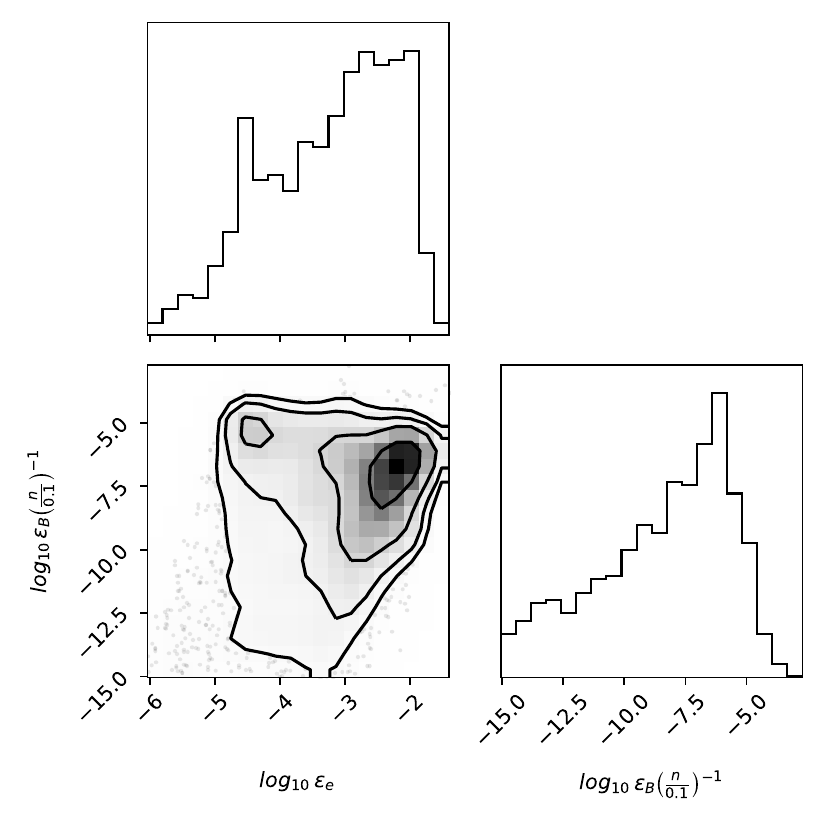}     
	\label{fig:cornerSSC2}
\end{figure}

To access the effects of the EIC cooling, we inject MeV photons to the comoving frame of the forward shock. The observed EIC radiation is significantly delayed with respect to the injection time. Moreover, at the time of EIC cooling, the jet/shock could be in the coasting, deceleration, or post-deceleration phase. It would depend on the initial bulk Lorentz factor of the jet, the density profile, and the density of the medium. Therefore, we restricted the injection of MeV photons at $\sim 250$ s (73 s considering the reference time of 177 s). We then compare the EIC spectrum at the time bin of $\sim 290$ s (112 s considering the reference time of 177 s). The electron population, the magnetic field, and the bulk Lorentz factor are rescaled from the late-epoch observations (1000-1350 s) following the self-similar solution of the relativistic shock in the homogeneous medium \cite{BM1976}. We do not allow for any free parameters beyond the modeling of 1000-1350 s SSC spectrum of the afterglow radiation. As can be seen in Fig. \ref{fig:Extended_model2}, the external injection of MeV photons, once corrected for the anisotropy, better describes the spectral shape of the early VHE afterglow of GRB 221009A. Note that our calculation of the afterglow spectrum with the prompt emission injection is only for the order of magnitude estimate. More realistic computations would require injection of time dependent spectra of the prompt emission, following both latitude-dependent contribution of pair-suppressed TeV emission and the delayed EIC radiation. Presence of additional features in the prompt emission spectrum, such as a low-energy hardening and high-energy softening or presence of the prompt VHE component, would further complicate the model. During the early epochs, also the intrinsic prompt emission spectrum could have been modified \cite{Beloborodov2002}, requiring more degrees of freedom in the model. 

\clearpage 

\bmhead{Other studies on the GRB 221009A afterglow radiation}

Several authors have investigated the multi-wavelength afterglow emission at late epochs (hours after the GRB trigger) and discussed the effects of the jet structure \cite{OConnor2023,Sato2023,Fraija2024,Laskar2023,Ren2024}, the radiation from the reverse shock \cite{Bright2023,Laskar2023}, the frequency-dependent growth of the apparent size of the afterglow emission region \cite{Giarratana2023}. Deep observations at the latest epochs did not reveal a clear signature of the supernovae signal \cite{Fulton2023,Levan2023,Shrestha2023}. We focus on the early (first $\sim$20 min) keV-TeV spectra of GRB 221009A. Previous studies of the early GeV-TeV radiation of GRB 221009A were limited to the investigations of temporal properties \cite{Ren2024} or joint GeV-TeV spectra with the piled-up data of Fermi/LAT (240-248 s in \cite{Wang2023}) or the GeV data in the off-source position of the Fermi/LAT (326-900 s in \cite{Wang2023}). 

\begin{table*}[ht!] \caption{The best-fit model parameters as shown in Fig. \ref{fig:BIN14_model}. The priors  for these parameters (first six rows) are listed in the second column. The final column displays the posterior values derived from the best-fit (see Fig. \ref{fig:cornerSSC1}). For parameters that are not constrained, such as magnetic field (B) and $\gamma_m$, we provide 95\% upper limits. The final two rows present the inferred equipartition parameters, $\epsilon_{e}$ and $\epsilon_{\rm B}$, assuming a uniform medium with a density of $n=0.1/$\,cm$^{3}$. The 95\% upper limits are also provided for these two quantities.}
    \centering
    \begin{tabular}{lcc}\hline
        Parameters              & Priors   & Posteriors\\ \hline
        log$_{10}$(B) [G]        &  (-5; 2) &   $<-1.3$\\
        log$_{10}(\gamma_{m})$   &  (0; 5)  &   $<3.5$\\
        log$_{10}(\gamma_{max})$ &  (4; 8)  &   6.6$^{+0.4}_{-0.2}$\\
        log$_{10} \rm l_{e}$        &  (-7; -1)&   -3.8$^{+0.7}_{-0.5}$\\
        $p$                     &  (2; 3)  &   2.7$^{+0.1}_{-0.2}$\\
        log$_{10}\rm D$         &  (1; 4)  &   2.8$^{+0.5}_{-0.3}$\\ \hline
        log$_{10} (\epsilon_{e})$ &    --      &   $<-1.9$ \\
        log$_{10} (\epsilon_{\rm B})$ &  --        &   $<-4.8 \left( \frac{n}{0.1} \right)^{-1}$ \\ \hline
    \end{tabular}
    \label{tab:bestfitmodel}
\end{table*}

\clearpage

\section*{Declarations}
\begin{itemize}
\item Acknowledgement:
BB expresses gratitude to Dr. Francesco Viola for the provision of computing resources. BB acknowledges financial support from the Italian Ministry of University and Research (MUR) for the PRIN grant METE under contract no. 2020KB33TP. We acknowledge the ModIC 2024 workshop (\url{https://indico.gssi.it/event/606/}) for providing an interdisciplinary platform for fruitful discussion. The work of DDF, AAL, DSS, and AET is supported by the basic funding program of the Ioffe Institute FFUG-2024-0002. AET also acknowledges accordo ASI e INAF HERMES 2022-25-HH.0. The authors thank the Director and the Computing and Network Service of the Laboratori Nazionali del Gran Sasso (LNGS-INFN). This research used resources of the LNGS HPC cluster realized in the framework of Spoke 0 and Spoke 5 of the ICSC project - Centro Nazionale di Ricerca in High Performance Computing, Big Data and Quantum Computing, funded by the NextGenerationEU European initiative through the Italian Ministry of University and Research, PNRR Mission 4, Component 2: Investment 1.4, Project code CN00000013 - CUP I53C21000340006. The authors thank fruitful discussion with Om Sharan Salafia, Maria Edvige Ravasio, Luca Foffano, Annalisa Celotti, Giancarlo Ghirlanda, Elena Amato, and Ulyana Dupletsa. 
\item Conflict of interest:
The authors declare that they have no conflict of interest. 
\item Author information statement: Correspondence and material requests should be directed to Biswajit Banerjee and Gor Oganesyan. 
\item Author contributions:
BB and GO originated the concept and spearheaded the project. BB ans SM led the LAT data reduction, analysis, and organization of data from AGILE and LHAASO, and authored the relevant sections of the article. AM, SM, ALDS and GO worked on the Fermi/GBM data reduction. ALDS has studied the MeV background, modified and implemented the pipelines for extraction of the background spectra. DF, AL, DS, and AT provided KONUS data and documented the data reduction process in the paper. JT and BB performed the log-parabola model fitting for the second component. GO and SM formulated the theoretical model to explain the observed data. SM and GO authored the modeling description in the paper. MB made significant contributions to the writing of the paper. BB and JT produced Fig. 1. BB worked out Figs. 2, 3, 7, 10, and 11. GO and BB created Figs. 4, 5, and 12. GO produced Figs. 6, 13, and 14. ALDS created Figs. 8 and 9. All authors participated in the discussions and editing of the paper. 

\end{itemize}


\bigskip








\clearpage 

\bibliographystyle{sn-standardnature}
\bibliography{references}
\end{document}


The Gamma-ray Burst Monitor (GBM) consists of 14 scintillating detectors, 12 NaI and 2 BGO. 

While GBM is not equipped with a veto to separate background and signal events, this is not typically problematic as GBM was designed to measure prompt GRB emissions. Characterized by a sharp peak above background. Given these fast events compared to the slower varying background, low order polynomials are often used to estimate the background around the considered signal region.


However, Over long time scales ($\sim ks$) and for relatively faint signals such as afterglow, polynomials do not necessarily accurately model the background. Mainly, the various components that contribute to the detector background have a strong variability over those time scales as proven by \cite{phys_model_gbm}.

To properly explain the background over such scales higher-order polynomials are necessary which introduces the risk of over-fitting the data.

An alternative for background estimation leverages rates from days before and after a time of interest. This is the concept introduced by \cite{fitzpatrick_osv} through the Fermi Orbital Subtraction Tool (OSV). 

The Fermi satellite passes over the same location on the Earth every $\sim$ 24 h, equivalent to every 15 orbits. Considering that the satellite rocks every two orbits to cover the whole sky, this means that every 30 orbits GBM has similar conditions both in position and orientation. By averaging the rates obtained 30 orbits ahead or before we obtain an estimate for the background for our time of interest. In \cite{fitzpatrick_osv} the authors also prescribe combining rates from 14 and 16 orbits before and after as they are near 15 orbits but with the correct orientation. Overall, they show how well this approach can approximate the background rates.

\begin{figure*}[h]\caption{Lightcurve for the BOAT for energies above 1 MeV, comparing the OSV estimated background with that of a polynomial fit}
\centering 
    \includegraphics[width=0.47\textwidth]{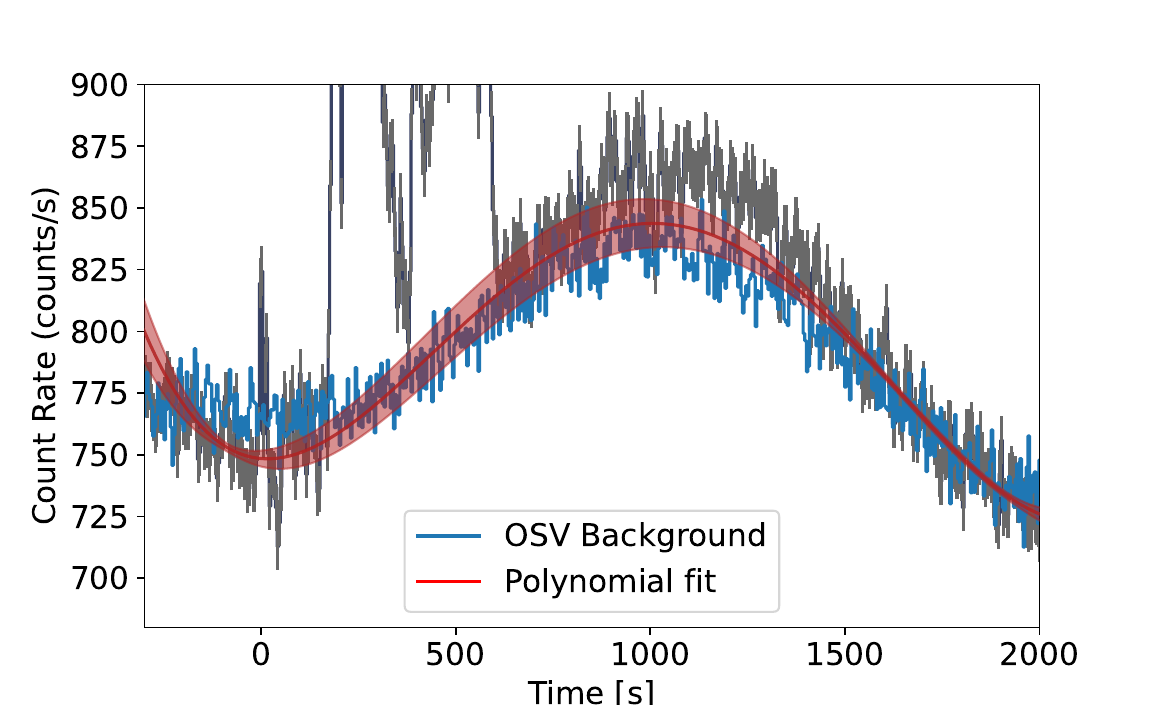}
    \includegraphics[width=0.47\textwidth]{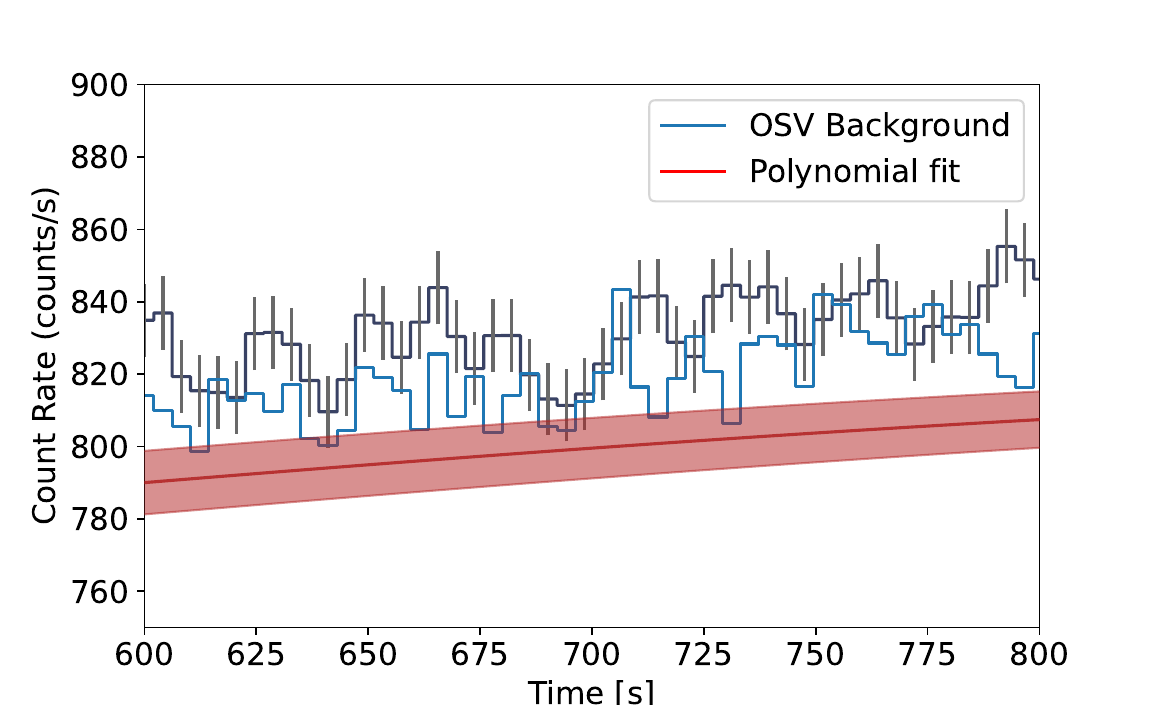}\label{appendix:fig:1MevCut}
\end{figure*}

We can directly illustrate the capabilities of this technique by applying it to the BOAT. Using the background intervals defined in \cite{BOAT_GBM_PAPER} to determine a polynomial fit, we can compare the two approaches. We can see in Fig.~\ref{appendix:fig:1MevCut}~(a) how the polynomial fit estimates a lower background than the OSV tool does. In the 600-800~s range (BIN 12 from Tab.~\ref{BIN_TABLE}) this corresponds to the difference between having a clear excess or having a signal compatible with the background (see Fig.~\ref{appendix:fig:1MevCut}~(b)).


\begin{figure}\caption{BOAT light curve with polynomial with considering the alternative cut described in Tab.~\ref{appendix:tab:new_intervals}}
    \centering
    \includegraphics[width=0.8\textwidth]{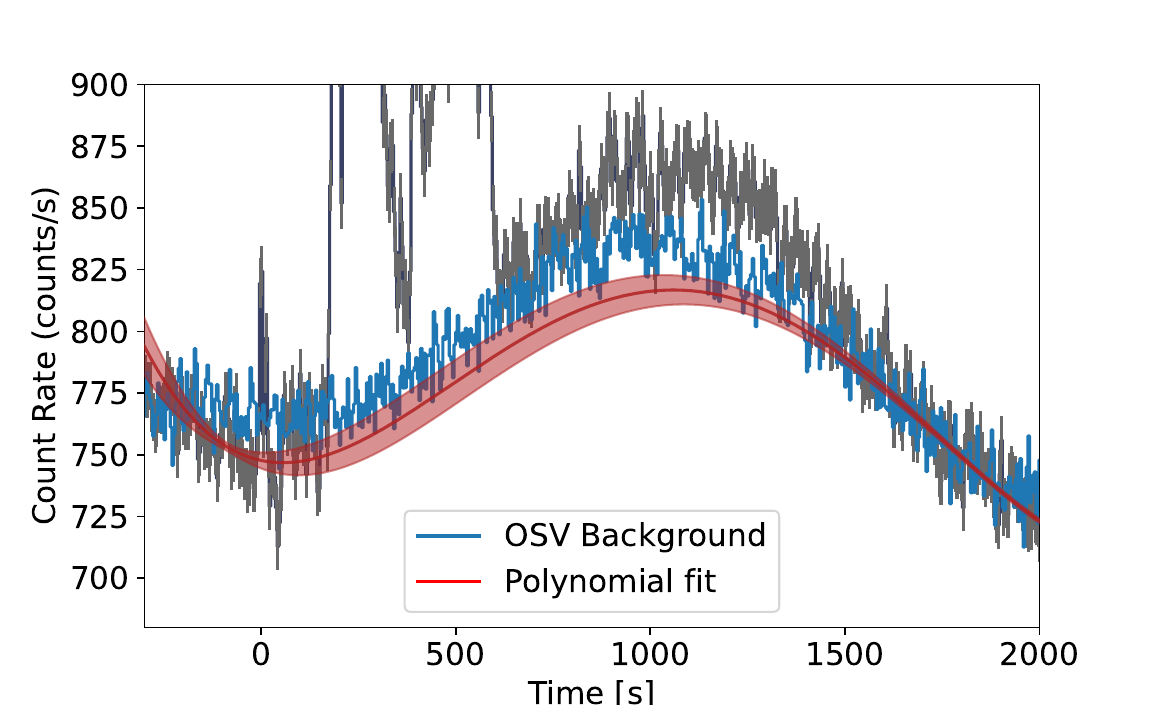}
    \label{appendix:fig:osv_fit_2}
\end{figure}

This is likely due to the polynomial fit being very sensitive to the background region we choose for fitting. In Tab.~\ref{appendix:tab:new_intervals} we show the (approximate) intervals used in \cite{BOAT_GBM_PAPER} and also an alternative selection of background ranges. The alternative selection takes into account that after 1470 the source is occluded by the Earth, allowing us to take a range closer to the actual signal. The resulting polynomial fit actually estimated an even higher background than the OSV tool, as can be seen in Fig.~\ref{appendix:fig:osv_fit_2}.

Given the fact that the OSV tool is less sensitive to the background region definition compared to a polynomial fit, we have shown a viable alternative for background estimation especially for long $\sim ks$ emissions.

The code used to generate the OSV background is based on the original code described in \cite{fitzpatrick_osv}. For the purpose of this paper some elements of the code were rewritten. The main difference is the use of Python 3 instead of Python 2, improving ease of installation (see \cite{De_Santis_2024}).

\begin{table}[ht]\caption{Two different possible selections for the background regions. Taking into account that the Earth occludes the source after 1470s.}
        \centering 
        \begin{tabular}{|c|c|c|}\hline
          Cut &  Left [s] & Right [s] \\ \hline
          Lesage et al. & (-150, -10) & (1625, 2400) \\
          Alternative & (-150, -10) & (1500, 2000) \\
          \hline
        \end{tabular}
        \label{appendix:tab:new_intervals}
\end{table}

\bibliographystyle{naturemag}
\bibliography{references}